\pdfoutput=1
\documentclass[11pt,twoside,a4paper,cmspaper,final,collab]{cms-tdr}
\def\svnVersion{a5b16c0}\def\svnDate{2026/02/24}\def\cmsCernNoTag{CERN-EP-2025-044}\def\cmsCernDate{\today}\def\cmsMessage{Published in the Journal of High Energy Physics as \href{http://dx.doi.org/10.1007/JHEP02(2026)141}{\doi{10.1007/JHEP02(2026)141}.}}
\begin{document}\cmsNoteHeader{EXO-23-010}

\newcommand{\tW}{\ensuremath{\PQt\PW}\xspace}
\newcommand{\WW}{\ensuremath{\PW\PW}\xspace}
\newcommand{\WZ}{\ensuremath{\PW\PZ}\xspace}
\newcommand{\ZZ}{\ensuremath{\PZ\PZ}\xspace}
\newcommand{\Nb}{\ensuremath{N_{\PQb}}\xspace}
\newcommand{\bsll}{\ensuremath{\PQb\PQs\Pell\Pell}\xspace}
\newcommand{\bbll}{\ensuremath{\PQb\PQb\Pell\Pell}\xspace}
\newcommand{\mll}{\ensuremath{m_{\Pell\Pell}}\xspace}
\newcommand{\Rmumuee}{\ensuremath{R_{\PGmp\PGmm/\Pep\Pem}}\xspace}
\newcommand{\lamdagstar}{\ensuremath{\Lambda/g_{*}}\xspace}
\providecommand{\cmsTable}[1]{\resizebox{\textwidth}{!}{#1}}

\cmsNoteHeader{EXO-23-010}
\title{Search for nonresonant new physics signals in high-mass dilepton events produced in association with \texorpdfstring{\PQb}{b}-tagged jets in proton-proton collisions at \texorpdfstring{$\sqrt{s} = 13\TeV$}{sqrt(s) = 13 TeV}}

\date{\today}

\abstract{
	A search for nonresonant new physics phenomena in high-mass dilepton events produced in association with \PQb-tagged jets is performed using proton-proton collision data collected in 2016--2018 by the CMS experiment at the CERN LHC, at a center-of-mass energy of 13\TeV corresponding to an integrated luminosity of 138\fbinv. The analysis considers two effective field theory models with dimension-six operators; involving four-fermion contact interactions between two leptons ($\Pell\Pell$, electrons or muons) and \PQb or \PQs quarks (\bbll and \bsll). Two lepton flavor combinations ($\Pe\Pe$ and $\PGm\PGm$) are required and events are classified as having 0, 1, or $\geq$2 \PQb-tagged jets in the final state. No significant excess is observed over the standard model backgrounds. Upper limits are set on the production cross section of the new physics signals. These translate into lower limits on the energy scale $\Lambda$ of 6.9 to 9.0\TeV in the \bbll model, depending on model parameters, and on the ratio of energy scale and effective coupling, \lamdagstar, of 2.0 to 2.6\TeV in the \bsll model. Lepton flavor universality is also tested by comparing the dielectron ($\Pe\Pe$) and dimuon ($\PGm\PGm$) mass spectra for different \PQb-tagged jet multiplicities. No significant deviation from the standard model expectation of unity is observed.}

\hypersetup{
  pdfauthor={CMS Collaboration},
  pdftitle={Search for nonresonant new physics signals in high-mass dilepton events produced in association with b-tagged jets in proton-proton collisions at sqrt(s)=13 TeV},
  pdfsubject={CMS},
  pdfkeywords={CMS, dilepton, b jets, BSM}}

\maketitle

\section{Introduction}

The standard model (SM) of particle physics successfully describes the known fundamental particles and the interactions between them. However, it fails to answer many open questions in particle physics, such as the nature of dark matter, the matter-antimatter asymmetry in the universe, the hierarchy problem, and the nonzero mass of neutrinos. To address the incompleteness of the SM, many theories beyond the SM have been introduced that predict the existence of new physics (NP) at the multi-\TeVns scale; a number of experiments at the CERN LHC are actively engaged in searches for NP signatures.

Searches for the resonant or nonresonant production of new physics at high invariant masses of dilepton events have played a central role in these programs. No significant sign of NP has been observed in these searches so far, with the latest results from the CMS and ATLAS Collaborations using proton-proton ($\Pp\Pp$) collision data at a center-of-mass energy $\sqrt{s} = 13\TeV$ collected during Run~2 of the LHC (2015--2018)~\cite{CMS:2021ctt,ATLAS:2019erb,ATLAS:2020yat}. Traditionally, these searches have mostly focused on final states inclusive in the number of \PQb-tagged jets. Recently, the ATLAS Collaboration published a search for nonresonant dilepton production in association with a \PQb-tagged jet using the Run~2 data~\cite{ATLAS:2021mla}, and the CMS Collaboration published a search for the resonant production of $\PGm\PGm$ pairs in association with \PQb-tagged jets with a similar data set, corresponding to an integrated luminosity of 138\fbinv~\cite{resonantAN}.

This paper presents the first search from CMS for new physics featuring the nonresonant production of dileptons in association with \PQb-tagged jets using the $\Pp\Pp$ collision data, collected from 2016 to 2018 and corresponding to an integrated luminosity of 138\fbinv. This search specifically considers the interaction of the NP state with \PQb quarks. The third-generation quark doublet is expected to be the most sensitive to NP because of the high mass of the quarks, and indeed hints of possible NP contributions have been given by experimental results on lepton flavor universality violation in \PB meson decays, notably in $R({\PDz})$ and $R({\PDst})$~\cite{LHCb:2023rd}. 

\begin{figure}[htp!]
  \centering
  \includegraphics[width=0.22\textwidth]{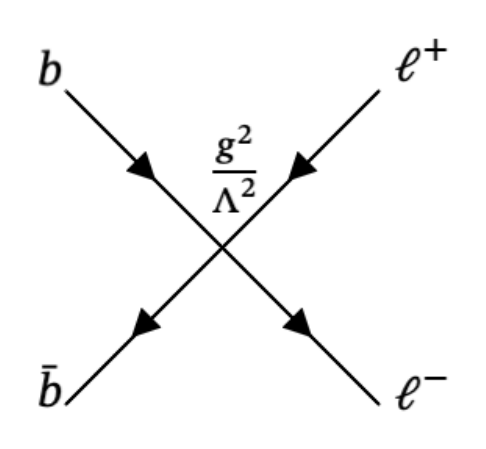}
  \includegraphics[width=0.33\textwidth]{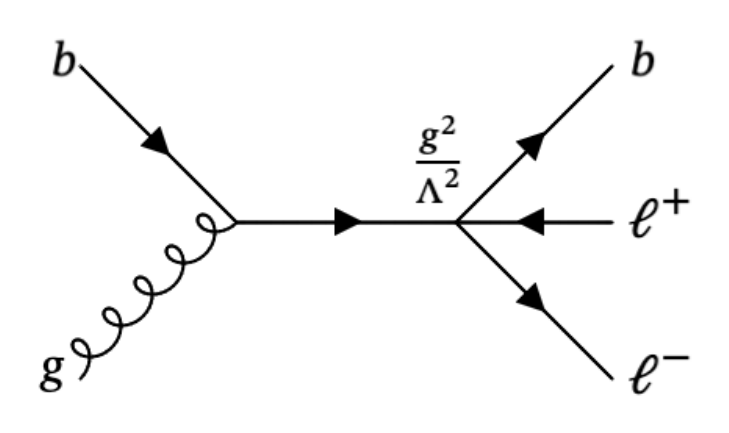}
  \includegraphics[width=0.28\textwidth]{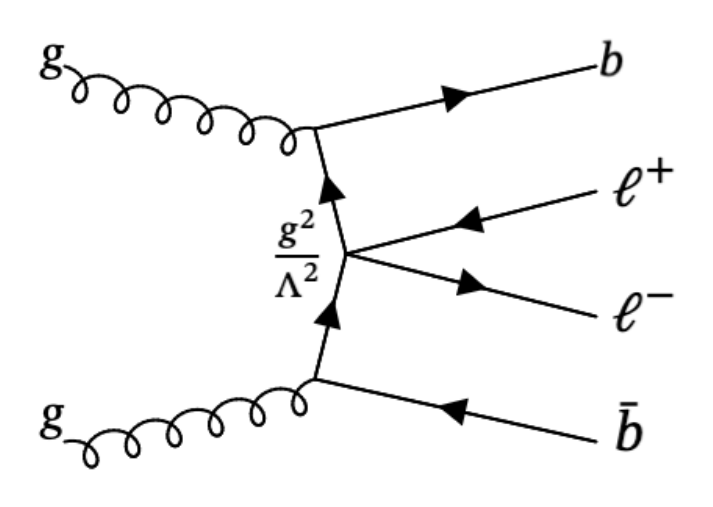}

  \caption{Representative Feynman diagrams for the production of dileptons via the $\PQb\PQb\Pell\Pell$ operator at the LHC, in association with 0 (left), 1 (center), and 2 (right) final state \PQb quarks.}

  \label{fig:bbll}
\end{figure}

The signal models considered in this analysis predict the nonresonant production of opposite-sign, same-flavor lepton pairs ($\Pep\Pem$, $\PGmp\PGmm$) in association with heavy-flavor jets. The nonresonant production of the NP can be modeled using effective field theory (EFT) techniques by integrating out the new degrees of freedom. Since the masses of the NP states associated with the signal are considered to be inaccessible at LHC energies, the models are described as a contact interaction (CI) between two leptons and two quarks ($\PQb\PAQb$ or $\PQb\PAQs$), resulting in a nonresonant excess of events at high mass. We target two such EFT models with dimension-six operators. One of the models is based on a pair of \PQb quarks and a pair of charged leptons (\bbll model), which can be generated by a tree-level exchange of a heavy neutral gauge boson \PZpr~\cite{Afik:2019htr} or scalar leptoquarks (LQs)~\cite{lq1,lq2,lq3,lq4}. Depending on how many of these \PQb quarks are sea quarks from within the colliding protons, the dilepton system is produced in association with 0 to 2 \PQb quarks in the final state. The analysis therefore targets different \PQb-tagged jet multiplicities (0, 1, and $\geq$2 \PQb-tagged jets). Figure~\ref{fig:bbll} shows the representative Feynman diagrams for the production of lepton pairs via the $\PQb\PQb\Pell\Pell$ operator at the LHC, in association with 0 (left), 1 (center), and 2 (right) final state \PQb quarks.

\begin{figure}[htp!]
  \centering
  \includegraphics[width=0.22\textwidth]{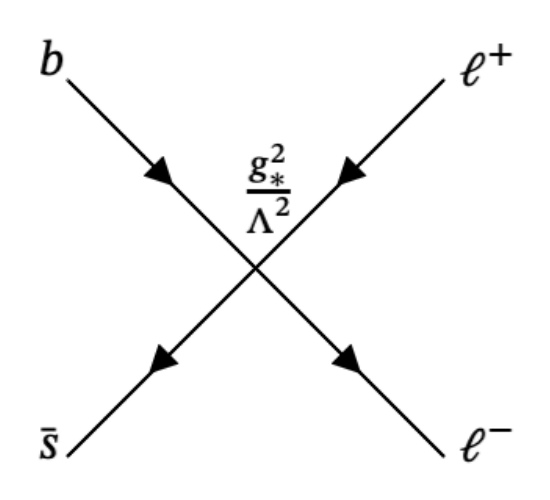}
  \includegraphics[width=0.30\textwidth]{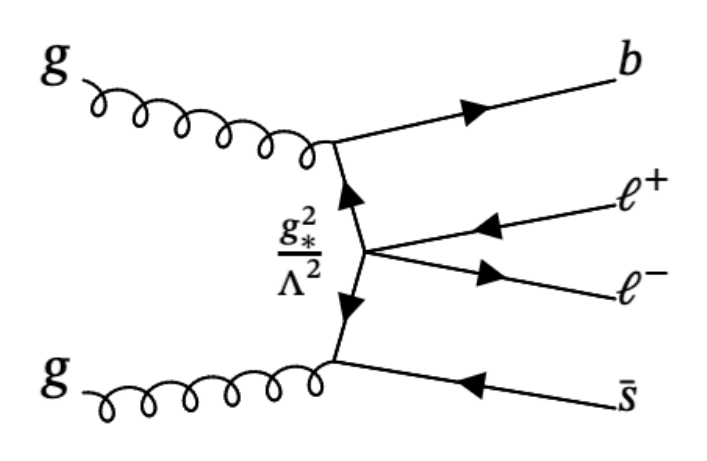}
  \caption{Representative Feynman diagrams for the production of dileptons via the $\PQb \PQs\Pell\Pell$ operator at the LHC, in association with 0 (left) and 1 (right) final state \PQb quarks.}

  \label{fig:bsll}
\end{figure}

The other model probed in this analysis considers the flavor-changing neutral current (FCNC) $\PQb\to \PQs\Pellp\Pellm$ transition (\bsll model), which is highly suppressed in the SM since it can only occur at loop level owing to the Glashow--Iliopoulos--Maiani (GIM) mechanism~\cite{Glashow:1970gm}. The $\PQb\to \PQs\Pellp\Pellm$ transition is expected to be mediated by a $\PZ^\prime$~\cite{Afik:2018nlr} or an LQ~\cite{bsll_lq1, bsll_lq2}, with flavor-violating couplings to $\PQb$ and $\PQs$ quarks, as well as couplings to either electrons or muons, as shown in Fig.~\ref{fig:bsll}. Any significant deviation from SM predictions in this channel would be a strong indication of beyond the SM physics. To target this model, $\Pe\Pe$ and $\Pgm\Pgm$ channels with exactly 0 and 1 $\PQb$-tagged jets are considered.

In addition to these two search strategies, lepton flavor universality is tested in these final states by measuring the $\PGm\PGm$-to-$\Pe\Pe$ ratio for different \PQb-tagged jet final states and comparing it to the corresponding SM expectation, after correcting for the differences in detector response, lepton acceptances, and lepton efficiencies. In contrast to the SM, the CI models predict a deviation from lepton universality~\cite{Greljo:2017vvb}, which would manifest itself in the $\PGm\PGm$-to-$\Pe\Pe$ differential cross section ratios.

Tabulated results are provided in the HEPData record for this analysis~\cite{hepdata}.

\section{The CMS detector}

The central feature of the CMS apparatus is a superconducting solenoid of 6\unit{m} internal diameter, providing a magnetic field of 3.8\unit{T} parallel to the proton beam axis. Within the solenoid volume are a silicon pixel and strip tracker, a lead tungstate crystal electromagnetic calorimeter (ECAL), and a brass and scintillator hadron calorimeter, each composed of a barrel and two endcap sections. Forward calorimeters extend the pseudorapidity ($\eta$) coverage provided by the barrel and endcap detectors. Muons are measured in gas-ionization detectors embedded in the steel flux-return yoke outside the solenoid. Events of interest are selected using a two-tiered trigger system. The first level (L1), composed of custom hardware processors, uses information from the calorimeters and muon detectors to select events at a rate of around 100\unit{kHz} within a fixed latency of 4\unit{$\mu$s}~\cite{Tapper:2013yva}. The second level, known as the high-level trigger (HLT), consists of a farm of processors running a version of the full event reconstruction software optimized for fast processing, and reduces the event rate to a few \unit{kHz} before data storage~\cite{Khachatryan:2016bia}. A more detailed description of the CMS detector, together with a definition of the coordinate system used and the relevant kinematic variables, can be found in Refs.~\cite{Chatrchyan:2008zzk, CMS-TDR-15-02}.

\section{Signal models}\label{sec:signal_model}
To probe new physics at the \TeVns scale, physics models based on EFT are considered that involve a four-fermion CI between a pair of leptons and a pair of quarks. This EFT approach involves dimension-six operators and limits the degrees of freedom related to heavy NP states by integrating them out. The introduction of CIs yields a modified SM Lagrangian with a series of higher-dimensional operators that are suppressed by inverse powers of the NP scale $\Lambda$:
\begin{equation}
  \mathcal{L}_{\text{eff}} = \mathcal{L}_{\text{SM}} + \frac{1}{\Lambda^2}\mathcal{L}_{2} + \ldots,
\end{equation}
where $\mathcal{L}_{\text{SM}}$ is the usual SM Lagrangian. The analysis focuses on two signal models based on this EFT approach. 

In the \bbll model, we consider the contribution of dimension-six operators to the $\PQq\PAQq \to \Pellp\Pellm$ process via a four-fermion CI between \PQb quarks and a pair of opposite-sign, same flavor leptons (electrons and muons). The effective Lagrangian for the $\PQb\PAQb\Pep\Pem$ and $\PQb\PAQb\PGmp\PGmm$ interaction can be written as:
\begin{equation}
  \mathcal{L}_{\text{eff}} = \mathcal{L}_{\text{SM}} + \frac{g^2}{\Lambda^2} \sum_{i,j=\text{L,R}} \eta_{ij}(\overline{b}_{i}\gamma_{\mu}b_{i})(\overline{\ell}_{j}\gamma^{\mu}\ell_{j}),
\end{equation}
where $\ell$ stands for \Pe or \PGm and $\Lambda$ is the scale of NP. The coupling strength $g$ is assumed to be $\sqrt{4\pi}$, following the convention in Ref.~\cite{Afik:2019htr}. The four-fermion CI consists of four possible chirality structures described by the $\eta_{ij} = {\pm}1$ term (left-handed quarks going to left-handed leptons (left-left, LL), left-handed quarks going to right-handed leptons (left-right, LR), etc.). By convention, only one of these possibilities is assumed to be realized in Nature at a time, setting the corresponding $\eta_{ij}$ to 1 and all others to zero. The sign of the $\eta_{ij}$ determines the sign of the interference of the signal with the SM Drell--Yan (DY) process, which can be either constructive ($\eta_{ij} = +1$) or destructive ($\eta_{ij} = -1$). This results in separate subclasses of this signal model, and all of them are considered in this analysis. It should be noted that the relative contribution from these CIs, compared to the SM DY process, grows with energy. For values of $\Lambda$ comparable to or smaller than the highest dilepton mass recorded at the LHC, the EFT expansion would diverge and lose validity. In this analysis, the lower limits on $\Lambda$ are set well above the maximum dilepton mass considered ($\approx$3.5\TeV), ensuring that the expansion safely converges in the probed parameter space~\cite{Busoni_2014}. 

The \bsll model considers the $\PQb\to \PQs\Pellp \Pellm$ transition with flavor-violating coupling to \PQb and \PQs quarks as well as couplings to either electrons or muons. The model is described by the effective Lagrangian
\begin{equation}
  \mathcal{L}_{\text{eff}} = \mathcal{L}_{\text{SM}} + \frac{C_{ij}^{U\Pell}}{v^2} (\overline{u}_\text{L}^i \gamma_\mu u_\text{L}^j)(\overline{\Pell}_\text{L} \gamma_\mu \Pell_\text{L}) + \frac{C_{ij}^{D\Pell}}{v^2} (\overline{d}_\text{L}^i \gamma_\mu d_\text{L}^j)(\overline{\Pell}_\text{L} \gamma_\mu \Pell_\text{L}),
\end{equation}

where $\mathcal{L}_{\text{SM}}$ denotes the SM Lagrangian, whose FCNC contributions are highly suppressed. The additional terms describe the EFT operators mediating transitions of up-type and down-type quarks to a lepton pair, respectively, with $v$ being the vacuum expectation value of the Higgs field. The flavor structure of the interaction is defined by the matrices $C_{ij}^{U\Pell}$ and $C_{ij}^{D\Pell}$, which are of the form:
\begin{equation}
  C_{ij}^{U\Pell} = \begin{pmatrix} 
    C_{\mathrm{u}\Pell} & 0 & 0 \\
    0 & C_{\mathrm{c}\Pell} & 0 \\
    0 & 0 & C_{\mathrm{t}\Pell} \\
  \end{pmatrix}, \,
  C_{ij}^{D\Pell} = \begin{pmatrix} 
    C_{\mathrm{d}\Pell} & 0 & 0 \\
    0 & C_{\mathrm{s}\Pell} & C^*_{\mathrm{bs}\Pell} \\
    0 & C_{\mathrm{bs}\Pell} & C_{\mathrm{b}\Pell}
  \end{pmatrix}.
\end{equation} 
Here, the $C_{\mathrm{bs}\Pell}$ term defines the new physics component and is given by the effective coupling ($g_*$) and energy scale ($\Lambda$) via $C_{\mathrm{bs}\Pell} = g_*^2v^2/\Lambda^2$, so that the ratio \lamdagstar is the free parameter of the model.

\section{Event simulation}\label{sec:samples}

{\tolerance=1800 Simulated Monte Carlo (MC) samples are generated for both the SM backgrounds and the new physics signals. The SM DY+jets, \WZ, \ZZ, and VVV~(V = \PW, \PZ) processes are simulated at next-to-leading order (NLO) in perturbative quantum chromodynamics using \MGvATNLO~v2.6.5~\cite{Alwall:2014hca}. The FxFx~\cite{Frederix:2012ps} scheme is used to match jets from matrix element calculations and parton showers. The DY+jets samples are produced binned in both the dilepton invariant mass (\mll) and the number of additional jet emissions, with a sample size large enough to ensure sufficient events at high \mll in different \PQb-tagged jets final states. The \ttbar, \tW, \WW, and SM Higgs boson processes are simulated at NLO using the \POWHEG~2.0~\cite{Nason:2004rx, powheg,Alioli_2010, Nason:2013ydw, Re:2010bp, Alioli:2008tz, Nason:2009ai} event generator. The diboson processes, including \WW, \WZ, and \ZZ, are referred to as VV processes.\par}

Event samples for both EFT models considered in the analysis are generated with different values for $\Lambda$. Events are generated at leading order (LO) by interfacing the \textsc{FeynRules} UFO~\cite{Degrande:2011ua,Afik:2019htr,Afik:2018nlr} model for \bbll (\bsll) with \MGvATNLO~v2.6.5 (v2.9.9). The MLM~\cite{Alwall:2007fs} scheme is used for jet matching. For the \bbll signal with constructive interference, samples are generated for $\Lambda$ values between 6 and 26\TeV in steps of 4\TeV, split into four exclusive bins in the dilepton invariant mass with lower bin edges at 300, 800, 1300, and 2000\GeV. For destructive interference, the samples are simulated with both NP and SM DY contributions to ensure net positive cross sections in these samples. These samples are generated with $\Lambda$ values between 4 and 18\TeV in steps of 2\TeV, split into four exclusive bins in the dilepton invariant mass with lower bin edges at 300, 800, 1300, and 2000\GeV. The samples are generated for all possible chirality structures (LL, LR, RL, RR) with $\eta_{ij} = \pm 1$, where the sign implies constructive ($+$1) or destructive ($-$1) interference with the SM DY process. For the \bsll model, samples are generated with \lamdagstar values ranging from 1 to 6\TeV in steps of 1\TeV, split into four exclusive bins in the dilepton invariant mass with lower bin edges at 200, 500, 1000, and 2000\GeV. 

The event generator outputs are interfaced with \PYTHIA~8.230~\cite{Sjostrand:2014zea} for parton showering and hadronization, using the \textsc{CP5} (\textsc{CP2}) tune~\cite{Sirunyan:2019dfx} for background (signal) samples. For all samples, the parton distribution functions (PDFs) are described with the NNPDF3.1 PDF set, computed at next-to-NLO~\cite{Ball:2017nwa} for background and LO for signal samples. The response of the CMS detector is simulated using the \GEANTfour~\cite{Agostinelli:2002hh} package. 

All samples are normalized to the integrated luminosity using the cross section calculated by the respective event generator. The presence of multiple $\Pp\Pp$ interactions in the same or adjacent bunch crossings (pileup) is incorporated by simulating additional interactions that are both in-time and out-of-time with the collision. The pileup in data is estimated based on the instantaneous luminosity measurements and the assumed inelastic $\Pp\Pp$ cross section. The simulated samples are reweighted such that the pileup profile matches the one observed in data, based on measurements of the luminosity and assuming that the $\Pp\Pp$ inelastic cross section is 69.2\unit{mb}~\cite{CMS:2020ebo}.

Because of the limited time resolution of the different subdetectors, objects in the L1 trigger can be assigned to the wrong bunch crossing, which can lead to a loss of events ("trigger prefiring"). While normally negligible, a slowly increasing shift of the reconstructed cluster time in the ECAL occurred in the L1 trigger during the 2016 and 2017 data taking, predominantly at high $\abs{\eta}$. The resulting efficiency loss is not included in simulation, and simulated events are reweighted to account for this inefficiency~\cite{CMS:2024ppo}. Similarly, a nonnegligible amount of trigger prefiring affected the L1 muon trigger during 2016 and to a lesser degree in 2017 and 2018, and this is taken into account in the analysis.

\section{Object reconstruction}
The reconstruction and selection of leptons in this analysis are largely unchanged from previous searches for new physics at large $m_{\Pell\Pell}$~\cite{CMS:2021ctt}. Compared to the previous results, the current analysis considers the \PQb-tagged jets in the final state to target different signal models, and the data have been reprocessed to account for updated calibrations of the detector.

A full reconstruction of the events is performed using a particle-flow (PF) algorithm~\cite{CMS:2017yfk}, which is based on an optimized combination of information from all subdetectors. The result is a list of PF candidates (electrons, muons, photons, as well as charged and neutral hadrons). The primary vertex is taken to be the vertex corresponding to the highest transverse momentum (\pt) scattering in the event, evaluated using tracking information alone, as described in Section 9.4.1 of Ref.~\cite{CMS-TDR-15-02}.

To be selected in the L1 trigger, events in the $\Pe\Pe$ channel must contain at least two electron candidates passing certain \pt thresholds. The thresholds evolved with time but never exceeded 25 (17)\GeV for the higher-(lower-)\pt electron. To protect against possible inefficiencies of the L1 trigger for electrons, events containing at least one electron, jet, or tau candidate passing significantly higher \pt thresholds at L1 are also included at the HLT level. In the HLT, the presence of two electrons with $\pt>33~(25)\GeV$ was required in 2016 and 2017 (2018).

Electron candidates are reconstructed from energy clusters in the ECAL, which are combined with matching tracks in the tracking system~\cite{CMS:2020uim}. The ECAL clustering groups together energy deposits that are compatible with having originated from the same particle, thus recovering energy losses due to bremsstrahlung. The angular information of the electron candidate is derived by combining information from the ECAL cluster and the associated track.

Electron candidates with $\pt>35\GeV$ and $\abs{\eta_{C}} < 1.44$ (ECAL barrel region) or $1.57 < \abs{\eta_{C}} < 2.50$ (ECAL endcap region) are selected, where $\eta_{C}$ is the pseudorapidity of the electron's ECAL cluster. As the quality of ECAL cluster reconstruction is lower in the transition region $1.44 < \abs{\eta_{C}} < 1.57$, electrons falling in that range are rejected. The electron candidates are required to pass selection and isolation criteria optimized for highly energetic electrons~\cite{CMS:2020uim}, which include a transverse impact parameter requirement of $\abs{d_\mathrm{xy}}<0.02\cm$ in the barrel and $\abs{d_\mathrm{xy}}<0.05\cm$ in the endcap, relative to the primary vertex. These criteria suppress backgrounds from pions, photon conversions, and electrons in jets. The electron efficiency scale factor (SF) is close to unity for both barrel and endcap, with an uncertainty ranging from 1--3\% across the \(\pt\) range of 35--1000\GeV, and set to 3\% above 1000\GeV.

The efficiency to trigger on, reconstruct, and identify high-\pt electrons has been studied in detail using $\PZ \to \Pe\Pe$ candidate events and has been validated in data up to \pt values of 1\TeV. In the analysis, simulated events are not required to pass the trigger selection, instead are weighted by the trigger efficiency measured in data. The trigger efficiency is observed to be $>$95\% for all data-taking years.

In the L1 trigger, events containing a single muon with $\pt>22\GeV$ are accepted. At the HLT, this threshold is 50\GeV throughout the whole data-taking period, while additional triggers using complementary reconstruction algorithms and a \pt threshold of 100\GeV were added in 2017 and 2018 to maximize efficiency for high-\pt muons. Muons within $\abs{\eta} < 2.4$ are accepted~\cite{CMS:2021yvr}.

Muons are reconstructed by combining the hits in the tracker and muon chambers. For muons with $\pt>200\GeV$, the TuneP algorithm is used to take into account radiative energy losses in the detector material~\cite{highpt_muons}. Muons are required to have $\pt>53\GeV$ and $\abs{\eta} < 2.4$. A set of selection criteria is applied that is optimized for high-\pt muons~\cite{highpt_muons}. Muons produced inside jets are rejected by requiring that the scalar sum of the \pt of all tracks not belonging to the muon candidate in a cone of $\Delta R = \sqrt{\smash[b]{(\Delta \eta)^2 + (\Delta \phi)^2}} < 0.3$ around the muon does not exceed 5\% of the muon's \pt. To further improve the selection purity of prompt muons, the muon track is required to have longitudinal distance $\abs{d_\mathrm{z}}<0.1\cm$, and transverse distance $\abs{d_\mathrm{xy}}<0.02\cm$ with respect to the primary vertex.

The efficiency of muons to pass the trigger, identification, and isolation requirements has been measured using $\PZ \to \Pgm\Pgm$ candidate events and validated up to 1\TeV. The efficiencies are stable as a function of \pt, with the exception of a slight loss of trigger efficiency at high \pt, caused by muons showering in the detector material and failing the L1 trigger. The SFs are observed to be close to unity, and are applied to the simulated events to account for small differences in efficiencies observed between data and simulation.

Using high-\pt muons originating from Lorentz-boosted \PZ bosons, the description of the muon momentum resolution in simulation has been compared with the data. Although good agreement is observed for central muons, the simulation underestimates the resolution effects for forward muons with $\abs{\eta} > 1.2$ by $\approx$15\%. The \pt values of these muons in simulated events have been smeared to account for this difference.

Jets are reconstructed from all PF candidates using the anti-\kt clustering algorithm~\cite{Cacciari:2008gp, Cacciari2} with a distance parameter of 0.4. Identification criteria are applied to jet candidates to remove anomalous effects from the calorimeters~\cite{CMS:2020ebo}. For jets with $\pt > 20\GeV$, the identification efficiency is $>$90\% depending on $\eta$. Jet energy scale (JES) corrections are applied to the measured jet energy depositions to correct for any offset due to energy coming from pileup interactions, miscalibration of the detector response to hadrons, and the residual differences between data and simulation as a function of the jet \pt and $\eta$. Jets are required to have $\pt>20\GeV$ and $\abs{\eta} < 2.4$. Jets originating from the hadronization of \PQb quarks are identified using the \textsc{DeepJet} algorithm~\cite{Bols_2020,CMS-DP-2023-005}. The \PQb quark jet with the highest \pt is required to satisfy the tight identification criteria, for which the identification efficiency is approximately 50\% (depending on \pt, $\eta$, and data-taking year), and the misidentification rate for light-flavor quark or gluon jets is 0.1\% and for \PQc quark jets is 2.4\%. The other \PQb quark jets in the event are required to satisfy relaxed requirements, for which the typical efficiency is 68\% (depending on \pt, $\eta$, and data-taking year), and the misidentification rate for light-flavor quark or gluon jets is 1.0\% and for \PQc quark jets is 12\%. The choice of tight and relaxed working points is made following the dedicated optimization studies within the CMS \PQb tagging framework, thus maximizing the overall signal significance across the different \PQb-tagged jet final states. The SFs are applied to correct for the difference in \PQb tagging performance between data and simulation~\cite{CMS:2017wtu}.

The missing transverse momentum \ptvecmiss is the negative vector \pt sum of all PF candidates~\cite{CMS:2019ctu}. Its magnitude is \ptmiss. The JES corrections are propagated to \ptmiss, which improves agreement in \ptmiss between simulation and data. Since the PF algorithm does not consider the dedicated high-\pt muon reconstruction using the TuneP algorithm, the \ptvecmiss is also corrected for the difference in muon \pt between the PF and TuneP algorithms.

\section{Event selection}
To search for a nonresonant excess of events in the dilepton mass spectra, electron or muon pairs are selected, where the leptons pass the selection described in the previous section. Since the charge misidentification probability increases significantly with electron \pt, resulting in a significant efficiency loss from an opposite-sign selection, no charge requirement is made for electron pairs. The fraction of same-sign events in DY simulation increases from about 1\% at the Z boson peak to $\approx$10\% for $\Pe\Pe$ masses of several \TeVns. In data, it is consistent with this expectation~\cite{CMS:2020uim, CMS:2021ctt}. In the case of muon pairs, the charge misidentification probability is of the order of $10^{-4}$ for high-\pt ($\approx$2\TeV) muons~\cite{highpt_muons}. Therefore, the two muons are required to have opposite electric charge. To suppress the contribution of muons resulting from cosmic rays, the angle between the two muons is required to be less than $\pi-0.02$. In the $\PGm\PGm$ channel, any event with an electron passing loose selection requirements with $\pt > 10\GeV$ and $\abs{\eta}<2.5$ is rejected, and in the electron channel events with a muon passing loose selection requirements with $\pt >10\GeV$ and $\abs{\eta}<2.4$ are rejected. 

For events containing more than two electrons or two muons ($<$1\%), originating mostly from \WZ and \ZZ production, the pair with an invariant mass \mll closest to the \PZ boson mass is considered. If no pair is found within 20\GeV of the nominal \PZ boson mass, the pair with the highest-\pt leptons is retained~\cite{CMS:2018ipm}. Finally, the search for NP is performed in events with $\mll>200~(300)\GeV$ for the \bsll (\bbll) signal and the events with $120 < \mll< 200\GeV$ are selected for a control region (CR) to estimate the DY+jets contribution.

In both the $\Pe\Pe$ and $\PGm\PGm$ channels, the events are divided into two categories based on the lepton $\abs{\eta}$~\cite{CMS:2021ctt}. This increases the sensitivity of the analysis by including the differences in lepton reconstruction performance and background levels between the central and forward parts of the detector. The ``barrel-barrel'' (BB) category is defined as both leptons having $\abs{\eta} < 1.44\,(1.20)$ for electrons (muons). The ``barrel-endcap'' (BE) category contains events where at least one electron (muon) has $\abs{\eta} > 1.56\,(1.20)$. Events with both leptons in the endcap (EE) are included in the BE category in the $\PGm\PGm$ channel, whereas such events in the $\Pe\Pe$ channel are rejected owing to a higher contribution of background in this channel. The events are then sorted into final state categories for zero \PQb-tagged jets (``0\PQb''), exactly one \PQb-tagged jet (``1\PQb''), and two or more \PQb-tagged jets (``2\PQb''). The \mll distributions are further divided into analysis bins according to the \PQb-tagged jet final states. Since the signal sensitivity depends on the binning, a dedicated study is performed to establish the optimal choice. For the 0\PQb and 1\PQb final states, the lower bin edges are set at 200, 300, 400, 500, 700, 1100, and 1900\GeV, with an overflow bin above 1900\GeV. For the 2\PQb final state, the lower bin edges are set at 200, 400, 600, and 900\GeV, with an overflow bin above 900\GeV. For the \bbll signal, the binning is the same as for the \bsll signal, except that the first bin starts at 300\GeV instead of 200\GeV. 

In the 1\PQb and 2\PQb final states, the dominant source of SM background arises from \ttbar production. To suppress this background, a deep neural network (DNN) based selection is used. The network is implemented using the \textsc{PyTorch} framework and trained separately for the $\Pe\Pe$ and $\PGm\PGm$ channels, and for each data-taking year, using variables where signal and background show significant differences, such as lepton and \PQb-tagged jet kinematic distributions, \ptmiss, the minimum invariant mass of a lepton and a \PQb-tagged jet, and the minimum 3D angle between the leading lepton and a \PQb-tagged jet. The network is trained using one representative \bbll signal sample ($\Lambda = 6\TeV$), and it was verified that the discriminating power does not degrade when it is applied to other signal models.

\begin{figure}[htb!]
  \centering
  \includegraphics[width=0.45\textwidth]{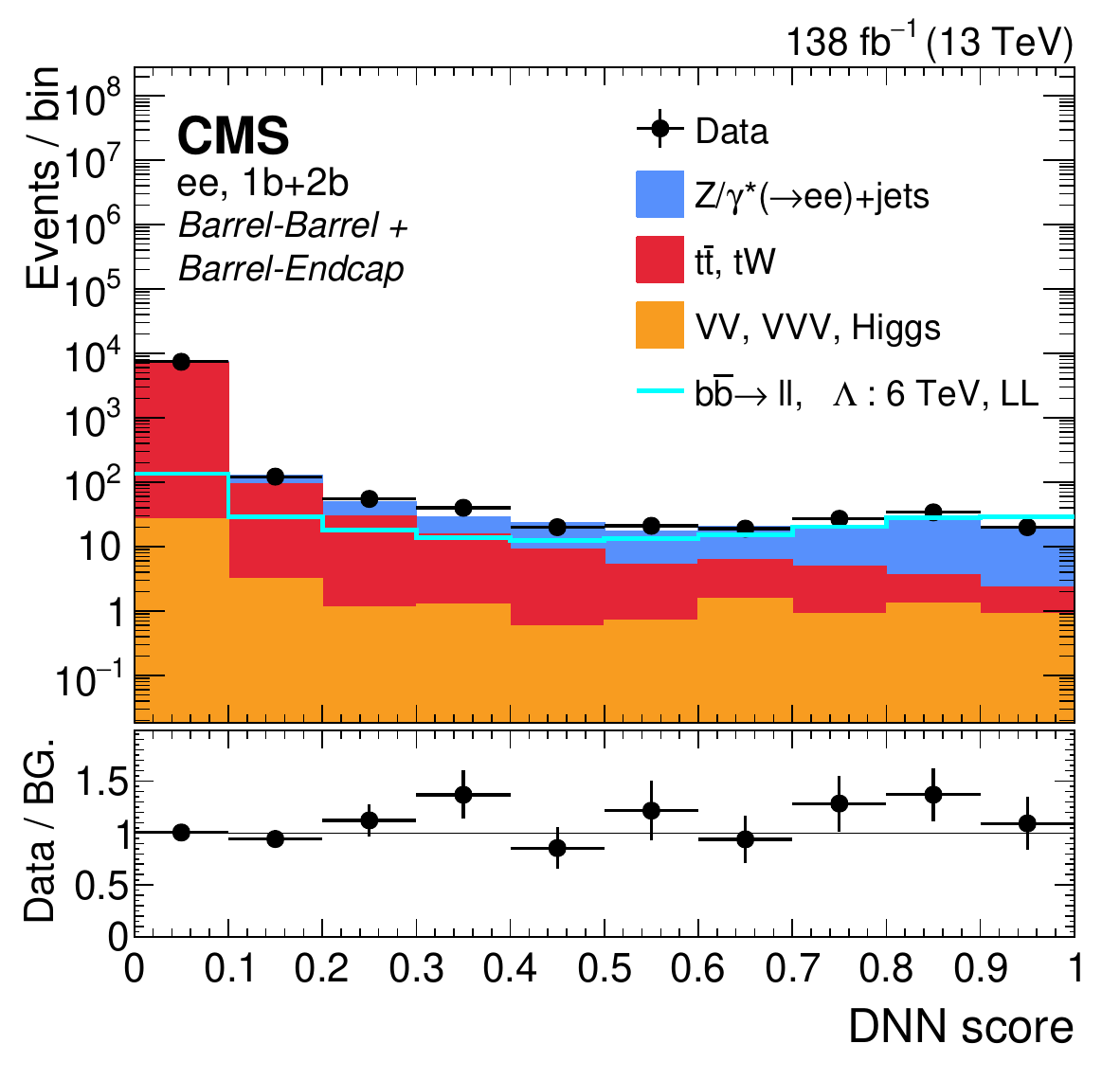}
  \includegraphics[width=0.45\textwidth]{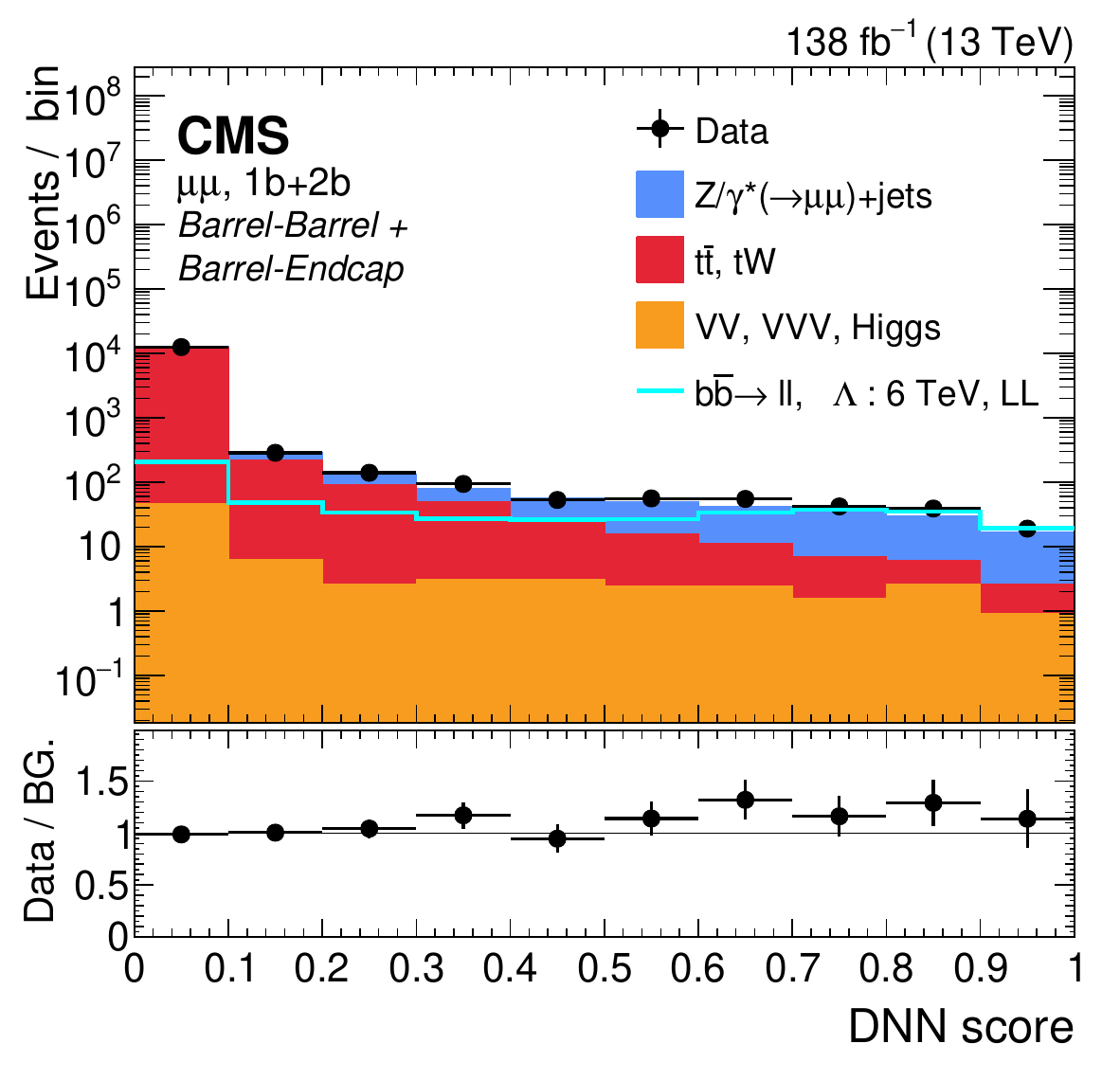}
  \caption{Observed data and various SM backgrounds in bins of DNN score evaluated from per-year and per-pseudorapidity trained models in the $\Pe\Pe$ (left) and $\PGm\PGm$ (right) channels. The solid cyan line in the upper panels correspond to the \bbll signal expectation for $\Lambda = 6\TeV$ in the LL constructive interference model. The lower panel of each plot shows the ratio of data to predicted background yields.}
  \label{fig:dnn_score_ee}
\end{figure}

The network is trained using the binary cross entropy loss classifier function, which measures the dissimilarity between true labels ($y_i = 1$ for signal and $y_i = 0$ for background) and the predicted probabilities for signal classification, optimizing the network to classify signal and background events effectively. For training the network, 80\% of this representative MC sample is selected, with the remaining 20\% used for validating the trained network. In addition, the final evaluation of the trained model is performed on independent data sets, drawn from different MC samples that are not used in training or validation, ensuring an unbiased performance assessment. The network is set to have four hidden layers with 128, 64, 32, and 16 nodes, respectively. The learning rate is varied at each epoch using an exponential decay scheduler and gradually decreases. The training is performed in 20 epochs, where it reaches the minimum and flat validation loss. This helps the model converge more effectively and provides a good balance between fast initial convergence and fine-tuning as the training progresses. Training and validation losses are found to be consistent, and the DNN score distributions for the training, validation, and testing data sets are in good agreement, confirming that overfitting is negligible. Figure~\ref{fig:dnn_score_ee} shows the DNN score, after combining models trained per-year, for data and SM backgrounds in the $\Pe\Pe$ (left) and $\PGm\PGm$ (right) channels. The plots are shown after applying the full signal region (SR) selections but before applying any scale factor corrections to the simulated \ttbar and DY+jets events. The contribution from nonprompt leptons is not shown, since it is negligible in the $\PGm\PGm$ channel and below 1\% in the $\Pe\Pe$ channel. The DNN score is required to be greater than 0.6 as this gives the best signal significance in simulation. Upon selecting events with this DNN score in the 1\PQb and 2\PQb final states, 90\% of \ttbar events are rejected; the DY+jets process becomes the dominant background in all event categories in the SR. 

\section{Background estimation}
Background processes producing electrons and muons are estimated from simulation. Dedicated control regions (CRs) are used to derive SFs that correct for possible mismodeling in simulation. The simulation of the dominant DY+jets background is corrected by comparing the simulated \mll spectra with those observed in data in the DY+jets CR, separately for $\Pe\Pe$ and $\Pgm\Pgm$ events. Before making this comparison, the \ttbar MC sample is corrected for any mismodeling of the selection efficiencies in simulation with an SF derived in the \ttbar CR ($\text{SF}_{\ttbar}$) as a function of the number of \PQb jets, \Nb, as described below. For the \ttbar process, additional validation regions (VRs) are used to verify the accuracy of these SFs before applying them in the SRs. The SR, CR and VR definitions are summarized in Table~\ref{tab:regions}.

\begin{table}[htb!]
  \centering
  \topcaption{Definitions of the SRs, CRs and VRs.}
  \begin{tabular}{lcccc}
    &  \mll range & Lepton flavor & DNN score  & \Nb \\ \hline
    0\PQb \bsll (\bbll) SR    &  $>$200 (300)\GeV & $\Pe\Pe$, $\Pgm\Pgm$  & \NA & $=$0\\
    1\PQb \bsll (\bbll) SR    &  $>$200 (300)\GeV & $\Pe\Pe$, $\Pgm\Pgm$  & $>$0.6 & $=$1\\
    2\PQb \bsll (\bbll) SR    &  $>$200 (300)\GeV & $\Pe\Pe$, $\Pgm\Pgm$  & $>$0.6 & $\geq$2\\

    DY CR   &  120--200\GeV & $\Pe\Pe$, $\Pgm\Pgm$ & \NA & $\geq$0 \\
    \ttbar CR    &  $>$200\GeV & $\Pe\Pgm$ & \NA & $\geq$0\\
    \ttbar VR    &  $>$200\GeV & $\Pe\Pgm$ & $>$0.6 & $\geq$1
  \end{tabular}
  \label{tab:regions}
\end{table}

\begin{figure}[htb!]
  \centering
  \includegraphics[width=0.45\textwidth]{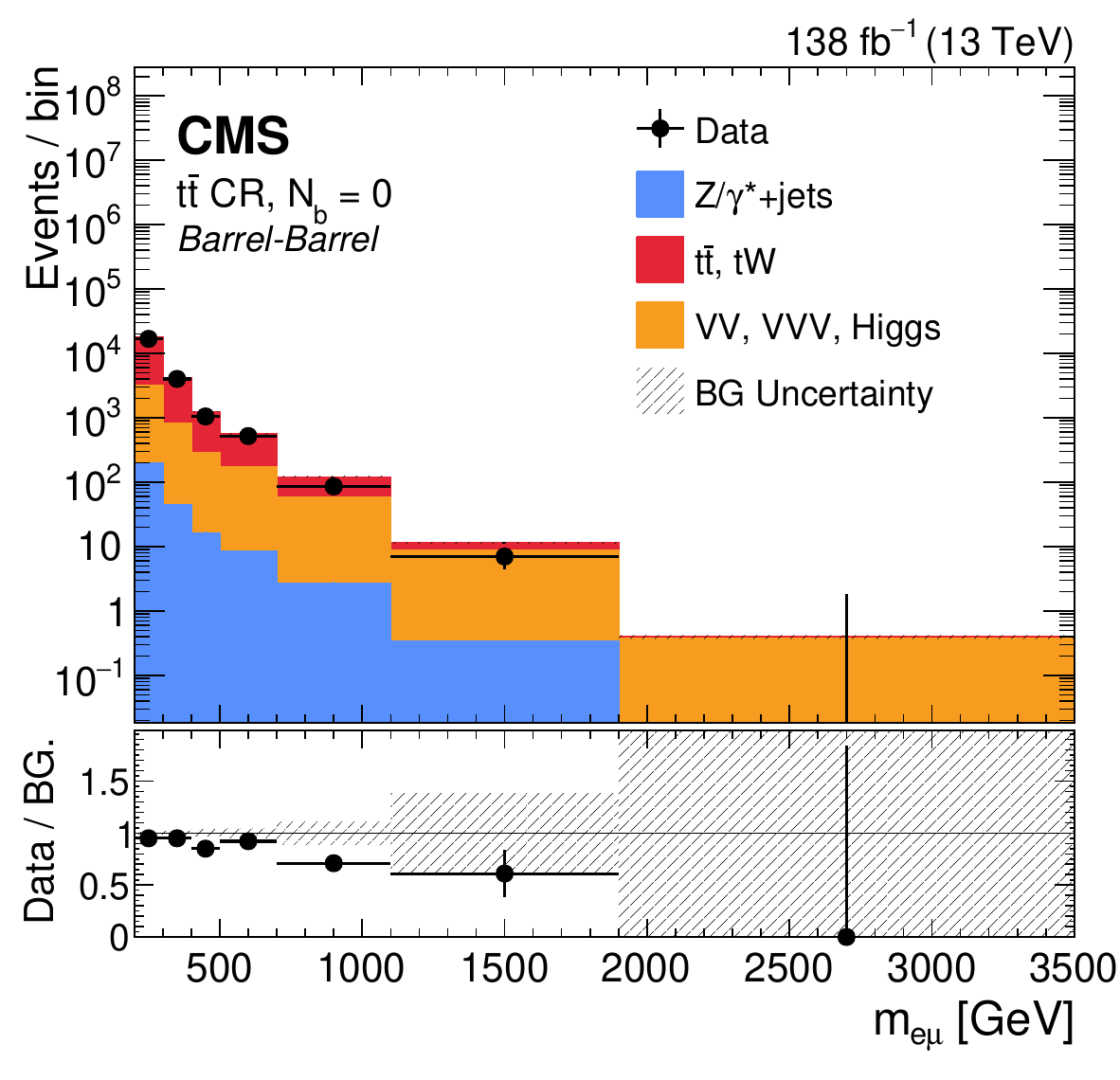}
  \includegraphics[width=0.45\textwidth]{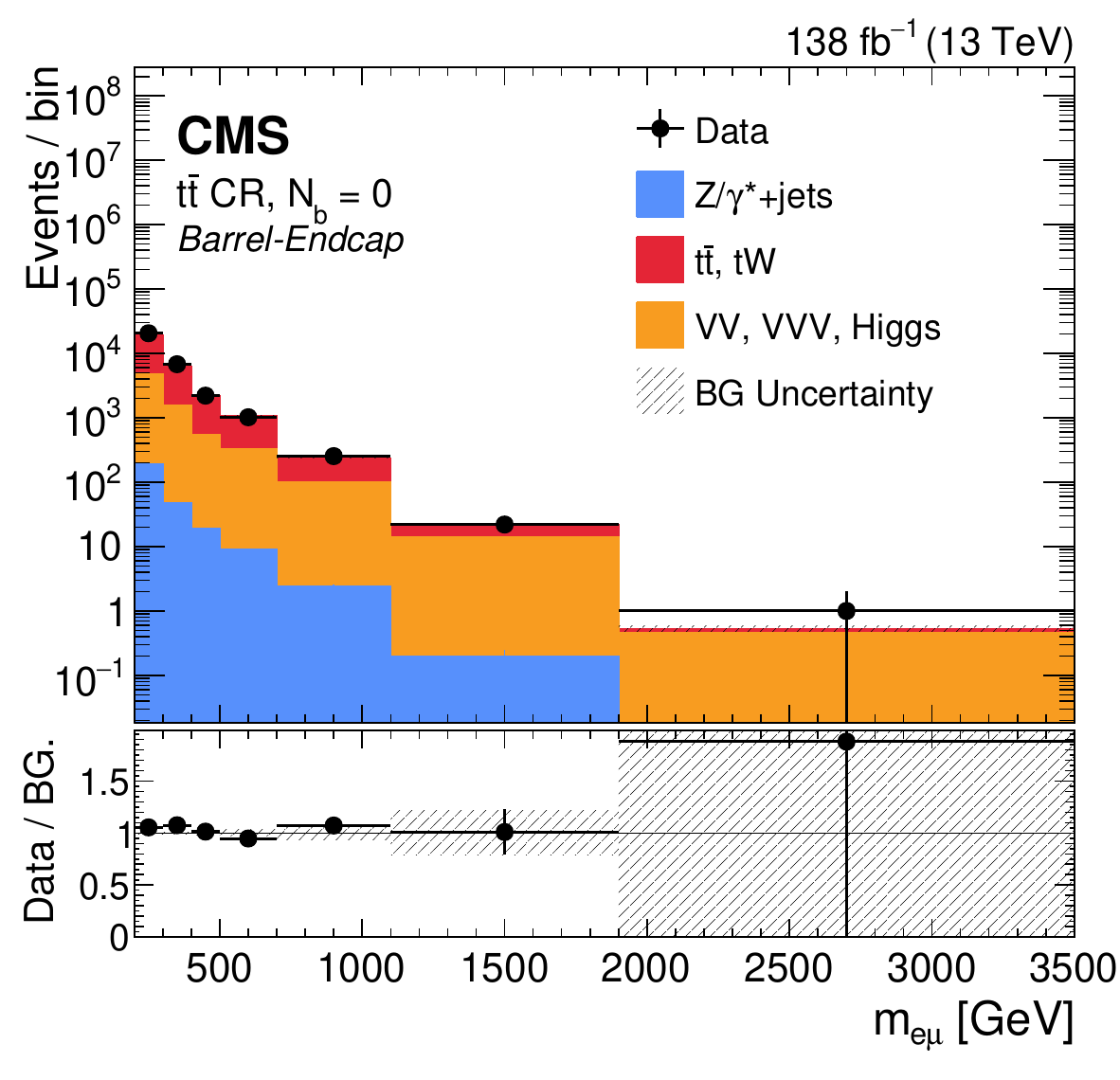}\\
  \includegraphics[width=0.45\textwidth]{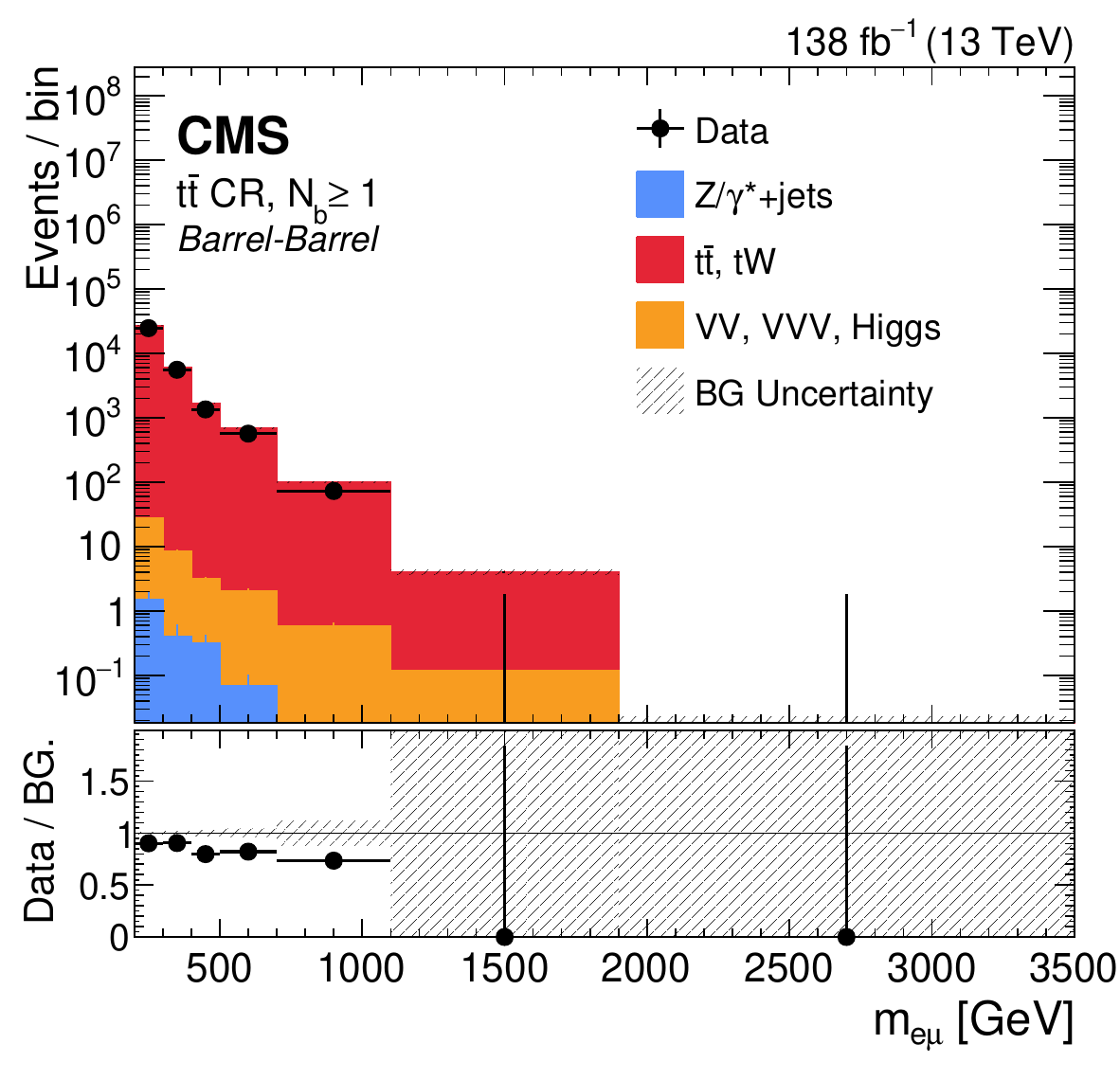}
  \includegraphics[width=0.45\textwidth]{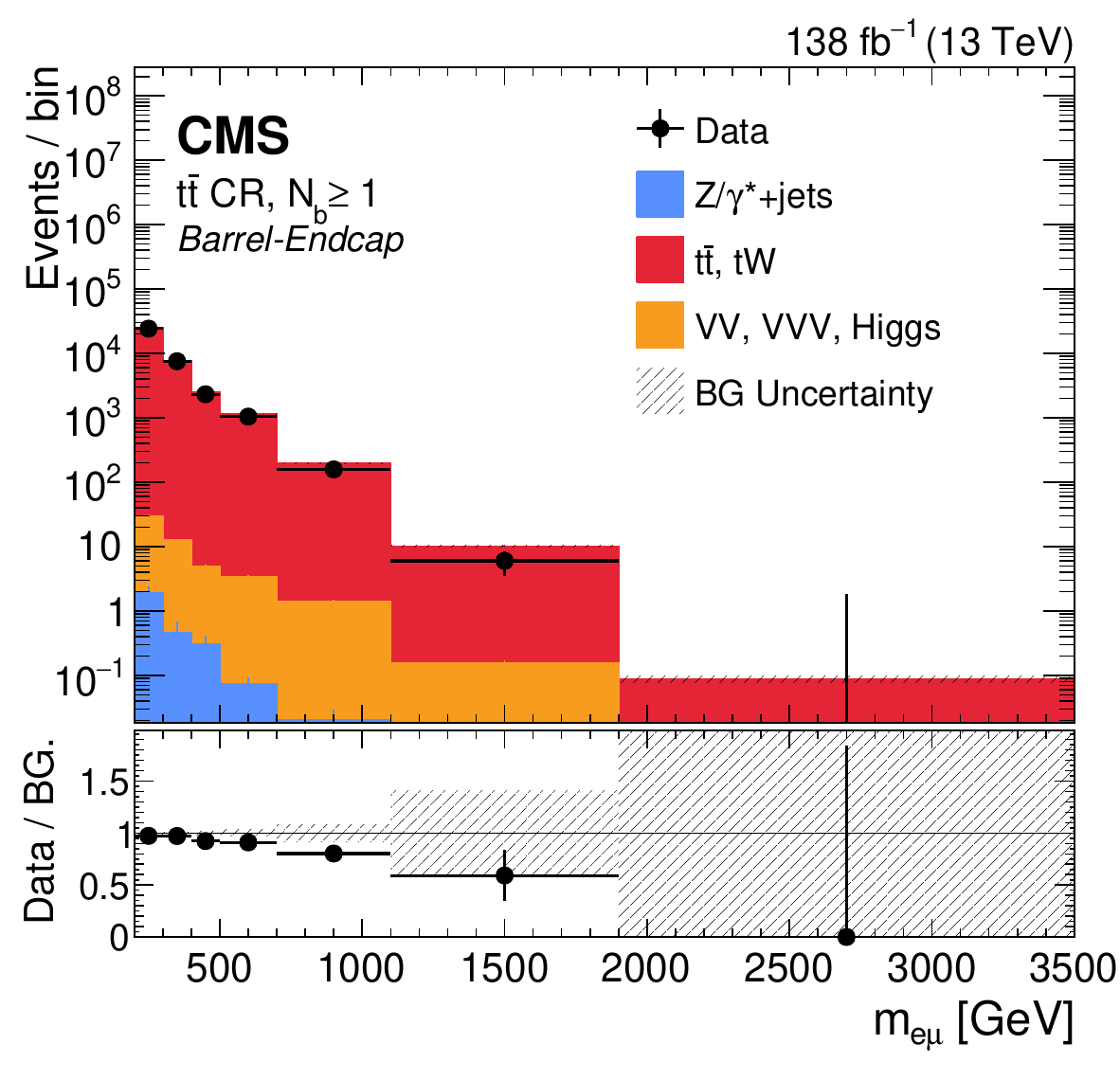}
  \caption{Comparison of data and various SM backgrounds in the \ttbar CR as a function of $m_{\Pe\PGm}$ for the 0\PQb (upper) and $\geq1$\PQb (lower) channels in the BB (left) and BE (right) categories. The lower panel of each plot shows the ratio of data to predicted background yields before correction with $\text{SF}_{\ttbar}$ values. The hatched band in the lower panels indicates the statistical uncertainty in the simulated background prediction.}
  \label{fig:ttbar_cr}
\end{figure}

The \ttbar CR is obtained by selecting an oppositely charged, different-flavor lepton pair ($\Pe\PGm$), where all the selection criteria for the $\Pe$ and $\PGm$ are kept the same as to those in the SR with $\mll > 200\GeV$ , except no DNN score selection is applied in this region. The resulting $\text{SF}_{\ttbar}$ is given by:
\begin{equation}
  \text{SF}_{\ttbar} = (N^{\text{Data}}_{\Pe\PGm} - N^{\text{non-\ttbar\, MC}}_{\Pe\PGm}) / N^{\ttbar\, \text{MC}}_{\Pe\PGm}.
\end{equation}
Here $N^{\text{Data}}_{\Pe\PGm}$ are the events observed in the \ttbar CR in data, and $N^{\text{non-\ttbar\, MC}}_{\Pe\PGm}$ and $N^{\ttbar\, \text{MC}}_{\Pe\PGm}$ are the events observed in the \ttbar CR in non-\ttbar MC samples and \ttbar MC samples, respectively. Because of the low event count in the $\Nb > 3$ category, the $\text{SF}_{\ttbar}$ estimated from the $\Nb = 3$ bin is used for the higher \Nb bins. The validation of simulation with data in the \ttbar CR is shown in Fig.~\ref{fig:ttbar_cr} as a function of $m_{\Pe\PGm}$ for the 0\PQb (upper) and $\geq$1\PQb (lower) channels. The purity of \ttbar events in the 0\PQb channel is around 80\%, whereas in the 1\PQb and 2\PQb channels, the purity is greater than 99\%.

The values of $\text{SF}_{\ttbar}$ measured in different bins of \PQb jet multiplicity are listed in Table~\ref{tab:sf_ttbar_bb} (\ref{tab:sf_ttbar_be}) for the BB (BE) category. The associated systematic uncertainties in the $\text{SF}_{\ttbar}$ are determined by propagating the statistical uncertainties in the data and simulation yields. 

\begin{table}[htp!]
  \centering
  \topcaption{Measured values of the $\text{SF}_{\ttbar}$ in the OS \ttbar CR (BB category).}
  \begin{tabular}{l c c c }
    \Nb  & 2016 & 2017 & 2018  \\ \hline

    0    & 0.92 $\pm$ 0.02   & 0.83 $\pm$ 0.01 & 0.84 $\pm$ 0.01 \\ 
    1    & 0.83 $\pm$ 0.01   & 0.83 $\pm$ 0.01 & 0.85 $\pm$ 0.01  \\ 
    2    & 0.84 $\pm$ 0.01   & 0.81 $\pm$ 0.01 & 0.83 $\pm$ 0.01  \\ 
    3    & 1.37 $\pm$ 0.10   & 1.15 $\pm$ 0.08 & 1.38 $\pm$ 0.07
  \end{tabular}
  \label{tab:sf_ttbar_bb}
\end{table}

\begin{table}[htp!]
  \centering
  \topcaption{Measured values of the $\text{SF}_{\ttbar}$ in the OS \ttbar CR (BE category).}
  \begin{tabular}{l c c c }
    \Nb  & 2016 & 2017 & 2018  \\ \hline
    0    & 1.10 $\pm$ 0.02 & 1.04 $\pm$ 0.01 & 1.00 $\pm$ 0.01 \\ 
    1    & 0.93 $\pm$ 0.01 & 0.93 $\pm$ 0.01 & 0.95 $\pm$ 0.01  \\ 
    2    & 0.94 $\pm$ 0.02 & 0.88 $\pm$ 0.01 & 0.91 $\pm$ 0.01 \\ 
    3    & 1.54 $\pm$ 0.11 & 1.26 $\pm$ 0.08 & 1.40 $\pm$ 0.07
  \end{tabular}
  \label{tab:sf_ttbar_be}
\end{table}

\begin{figure}[htb!]
  \centering
  \includegraphics[width=0.45\textwidth]{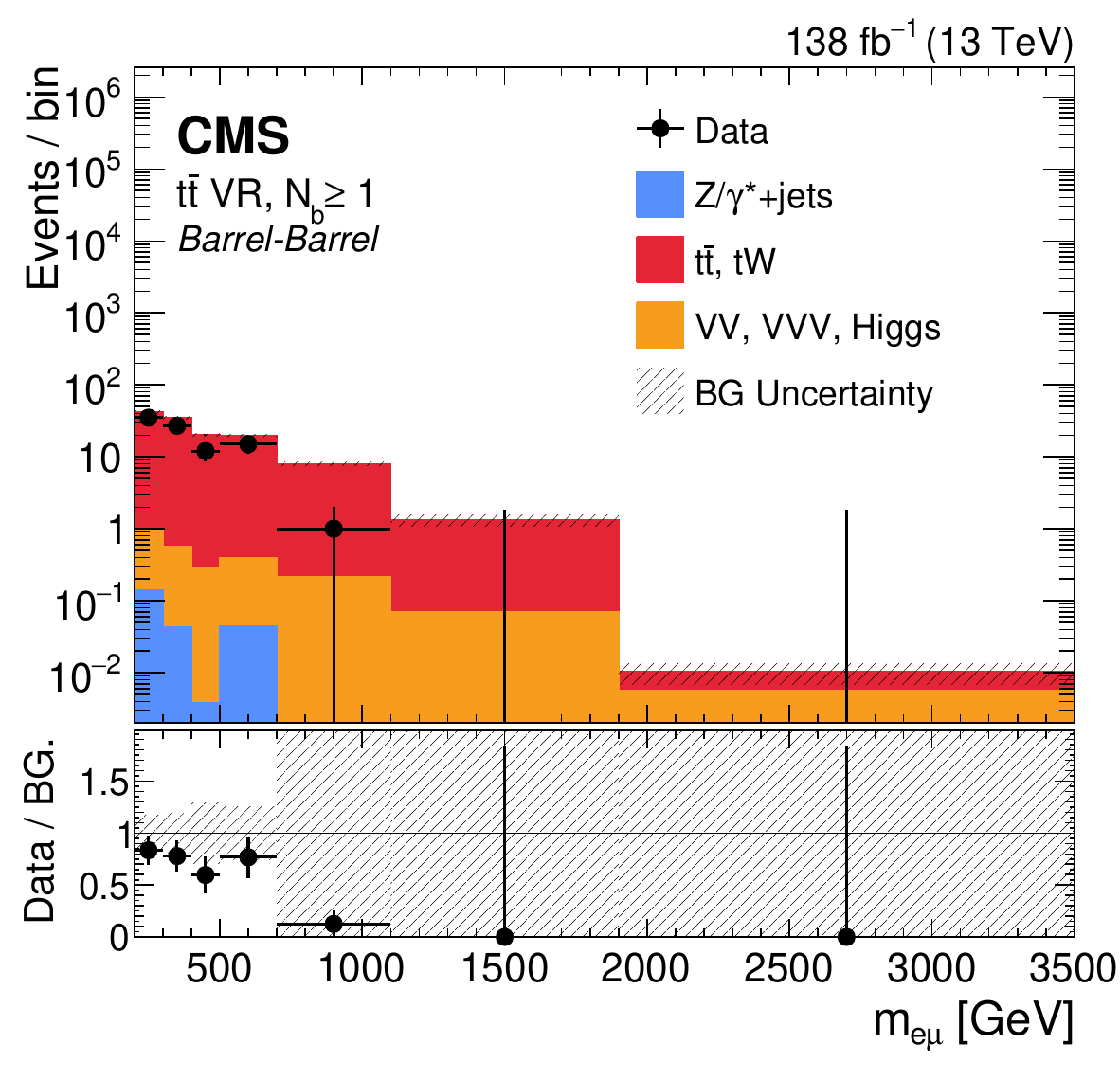}
  \includegraphics[width=0.45\textwidth]{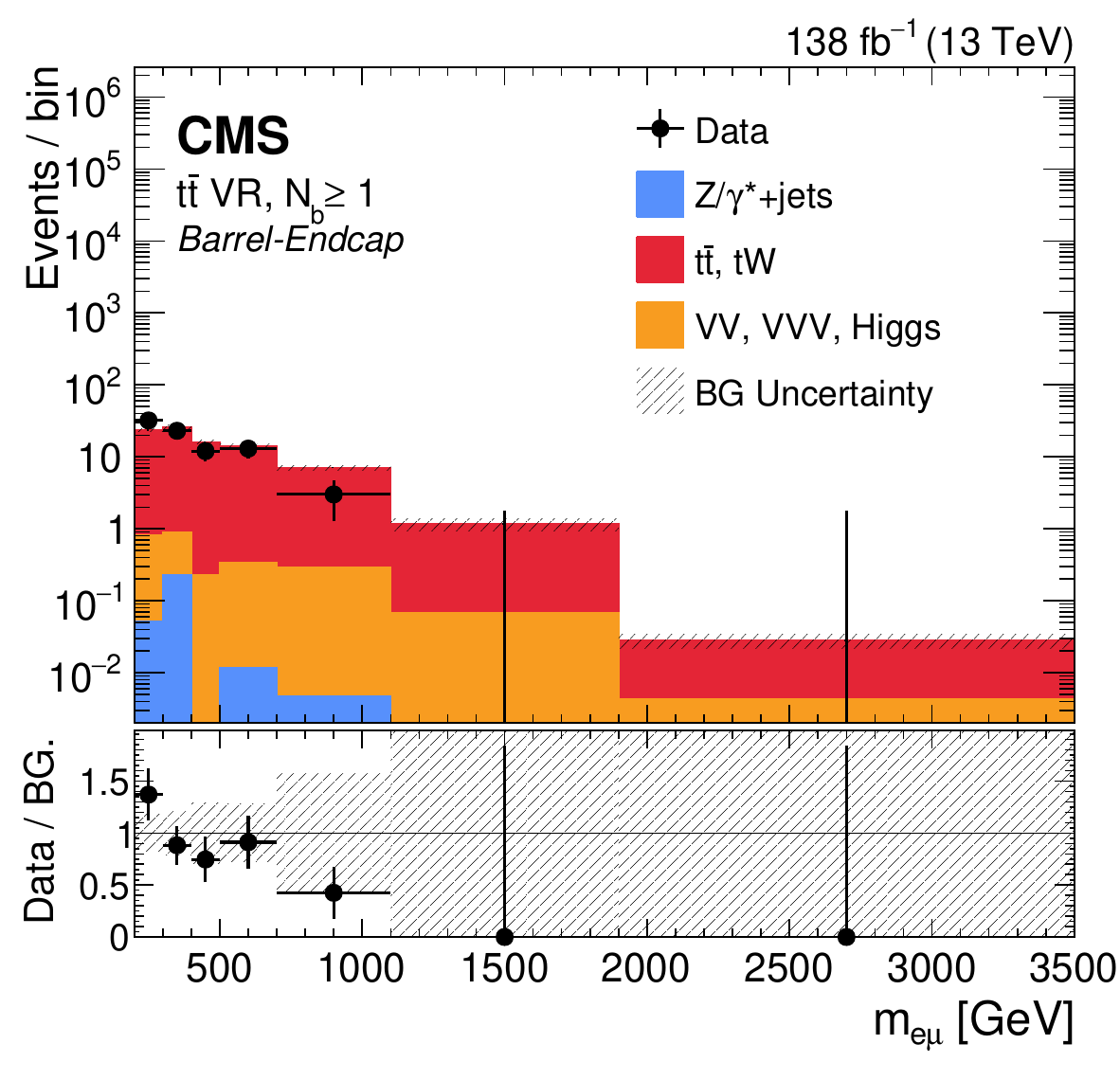}
  \caption{Comparison of data and various SM backgrounds in the \ttbar VR as a function of $m_{\Pe\PGm}$ for the BB (left) and BE (right) categories. The lower panel of each plot shows the ratio of data to predicted background yields. The hatched band in the lower panels indicates the statistical uncertainty in the simulated background prediction.}
  \label{fig:ttbar_val}
\end{figure}

Once the \ttbar events are corrected with this $\text{SF}_{\ttbar}$, a validation of observed \ttbar events in simulation is performed after the DNN score selection in the 1\PQb and 2\PQb channels. The \ttbar validation region (VR) is a subset of \ttbar CR with the additional requirements of DNN score to be greater than 0.6 and $\geq$1~\PQb-tagged jets. Figure~\ref{fig:ttbar_val} shows the comparison of data and various SM processes in the \ttbar VR after correcting with $\text{SF}_{\ttbar}$ values. The deviations from unity are taken to be the systematic uncertainty in the \ttbar estimation in the SR. 

\begin{figure}[htb!]
  \centering
  \includegraphics[width=0.45\textwidth]{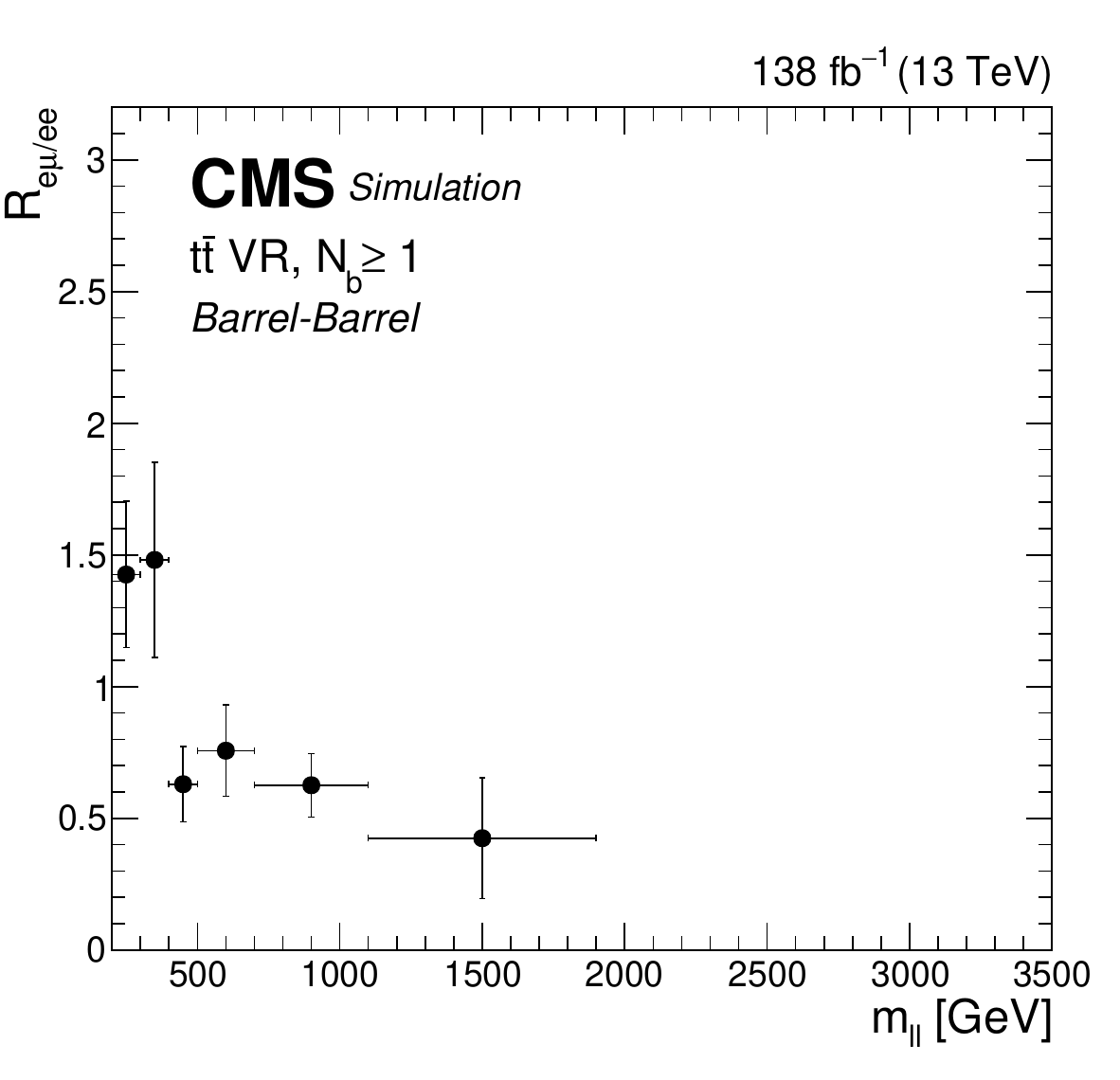}
  \includegraphics[width=0.45\textwidth]{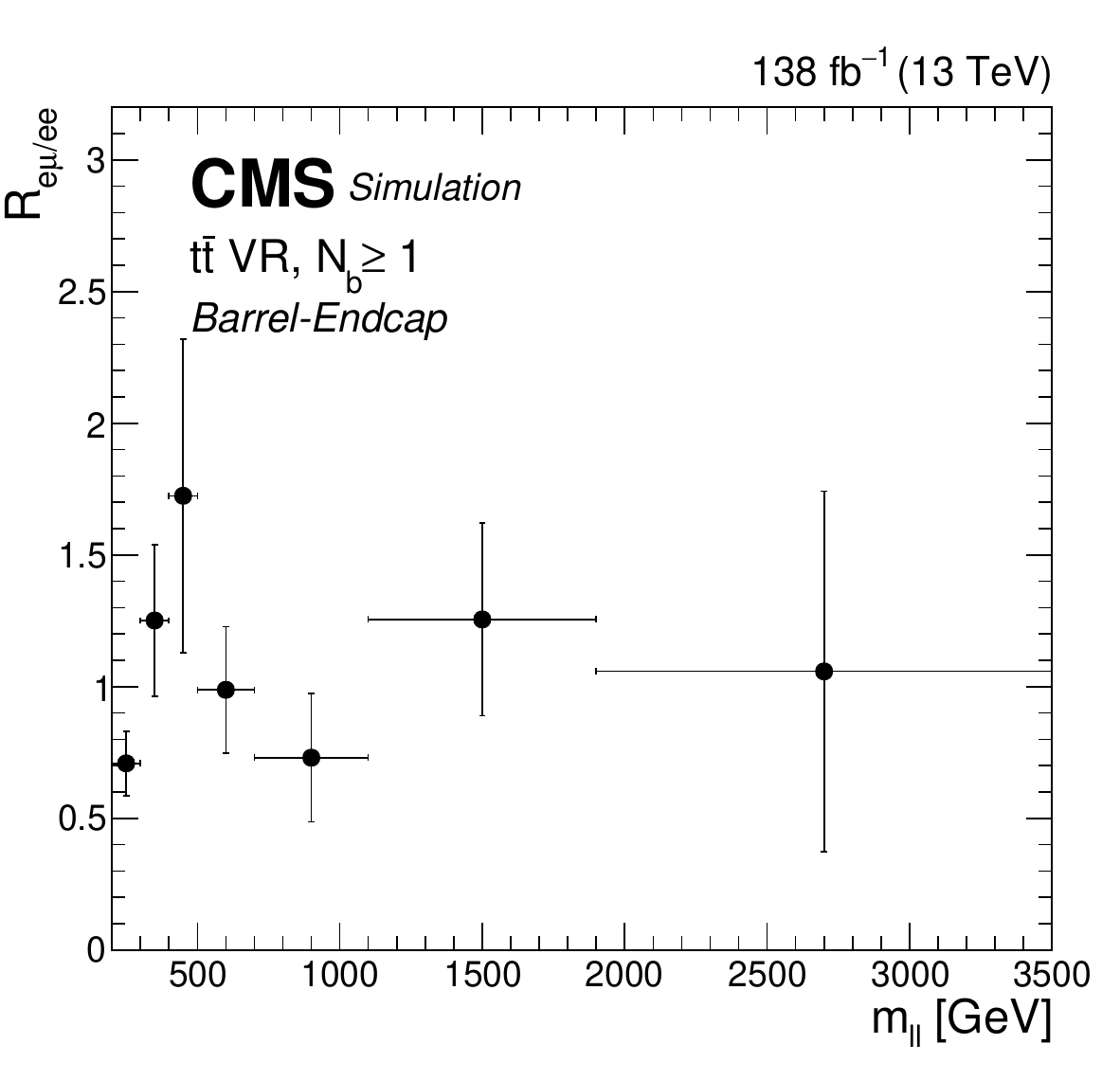}
  \caption{Ratio of simulated $\ttbar \to \Pe\PGm$ events with SR selections and simulated $\ttbar \to \Pe\Pe$ events in the SR for the BB (left) and BE (right) categories. The error bars represent the statistical uncertainty in the simulated event yields.}
  \label{fig:ttbar_shape_ee}
\end{figure}

\begin{figure}[htb!]
  \centering
  \includegraphics[width=0.45\textwidth]{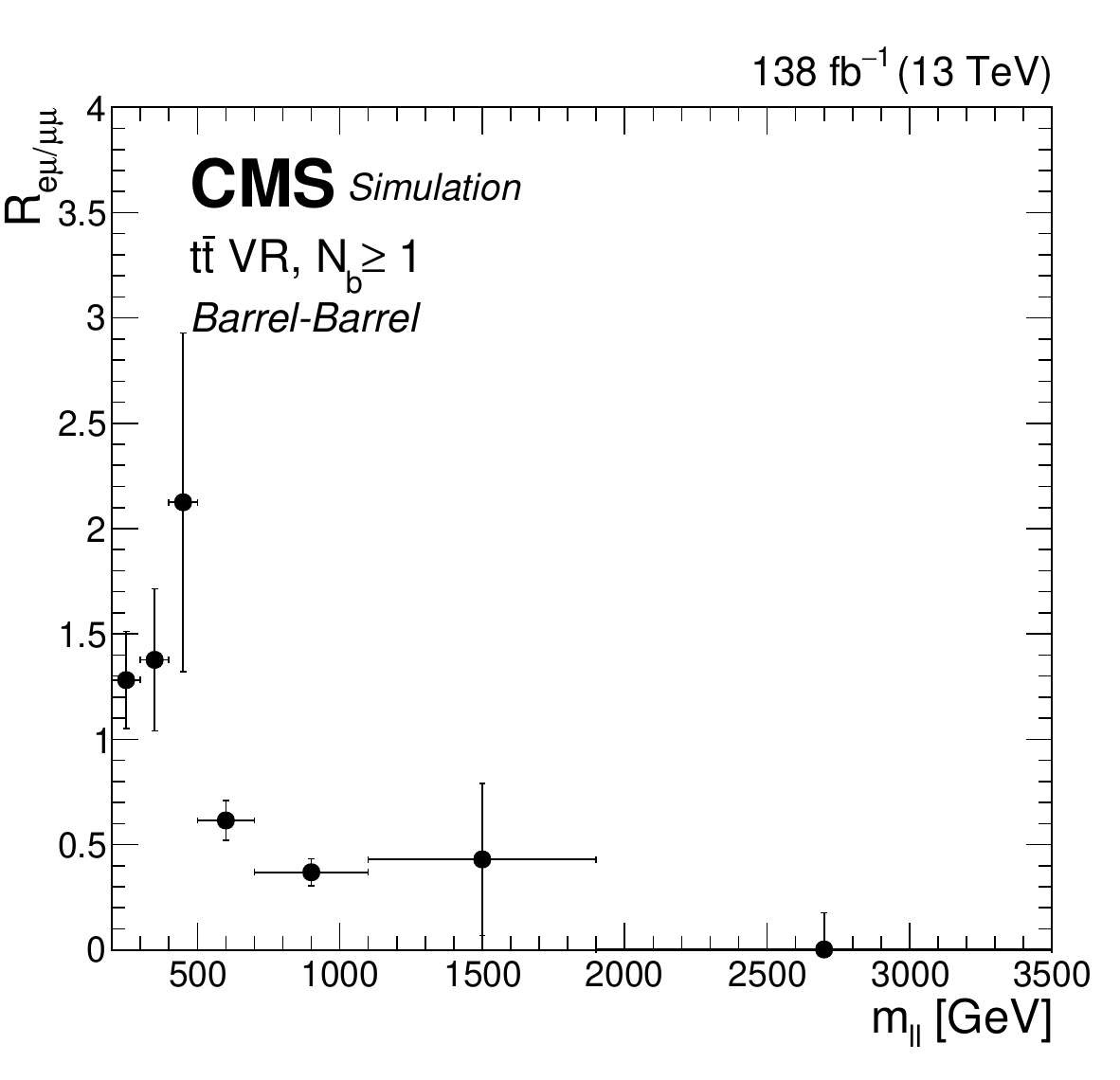}
  \includegraphics[width=0.45\textwidth]{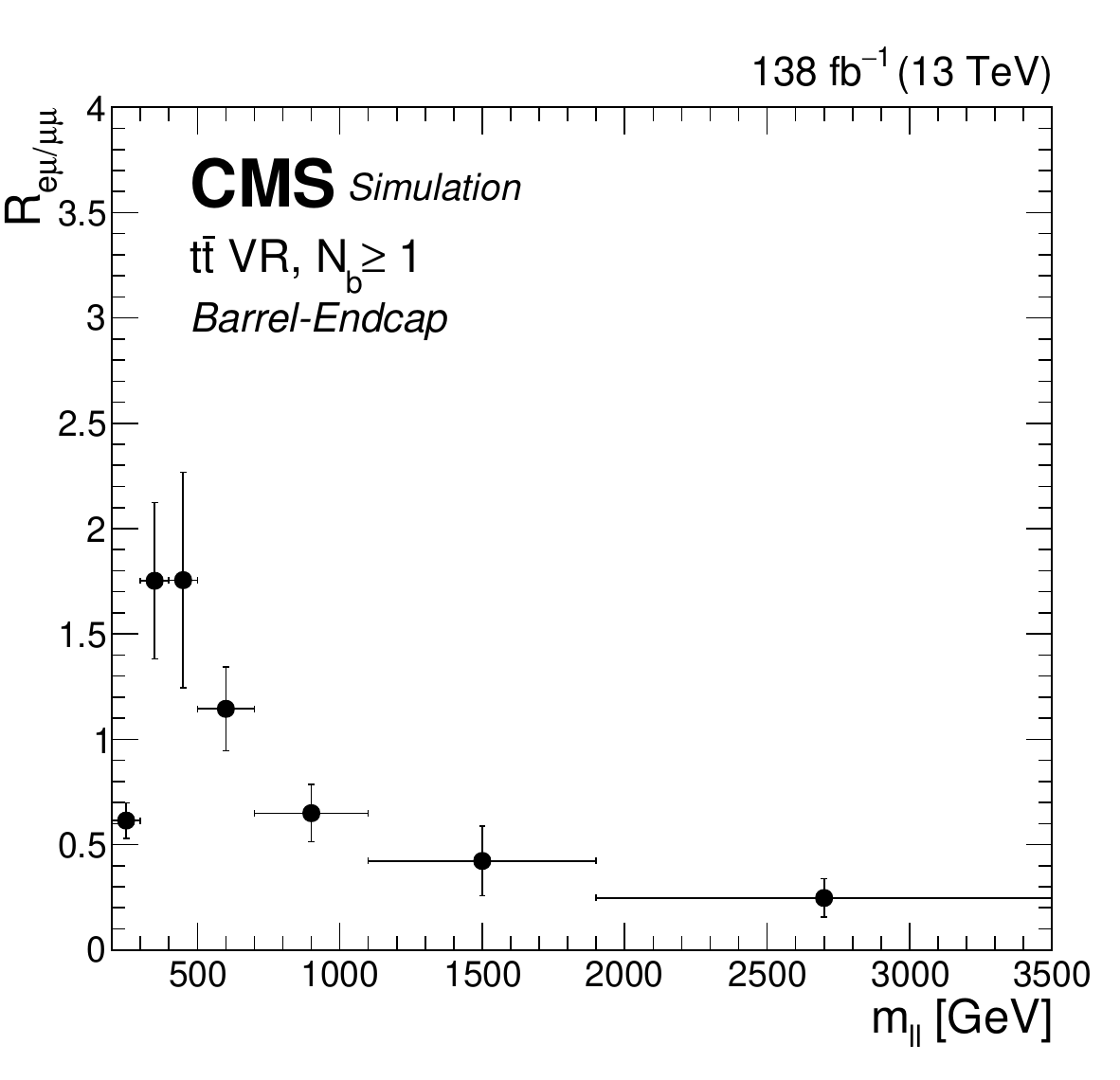}
  \caption{Ratio of simulated $\ttbar \to \Pe\PGm$ events with SR selections and simulated $\ttbar \to \PGm\PGm$ events in the SR for the BB (left) and BE (right) categories. The error bars represent the statistical uncertainty in the simulated event yields.}
  \label{fig:ttbar_shape_mumu}
\end{figure}

Additionally, the shapes of simulated \ttbar dilepton mass distributions as a function of $m_{\Pe\PGm}$ in the \ttbar VR and $m_{\Pe\Pe}$ ($m_{\PGm\PGm}$) in the SR are compared to ensure the robustness of the \ttbar modeling in the SR after the $\text{SF}_{\ttbar}$s are applied. The bin-by-bin deviations from unity in the ratio of the two distributions are considered as an additional systematic uncertainty, applied as a function of $m_{\Pe\Pe}$ ($m_{\PGm\PGm}$) bins, in the \ttbar background estimate in the SR. Figure~\ref{fig:ttbar_shape_ee} shows the ratio ($R_{\Pe\PGm/\Pe\Pe}$) of $m_{\Pe\PGm}$ and $m_{\Pe\Pe}$ distribution for the BB (left) and BE (right) categories. Similarly, Fig.~\ref{fig:ttbar_shape_mumu} shows the ratio ($R_{\Pe\PGm/\PGm\PGm}$) of $m_{\Pe\PGm}$ and $m_{\PGm\PGm}$ distribution for the BB (left) and BE (right) categories.

\begin{figure}[htb!]
  \centering
  \includegraphics[width=0.45\textwidth]{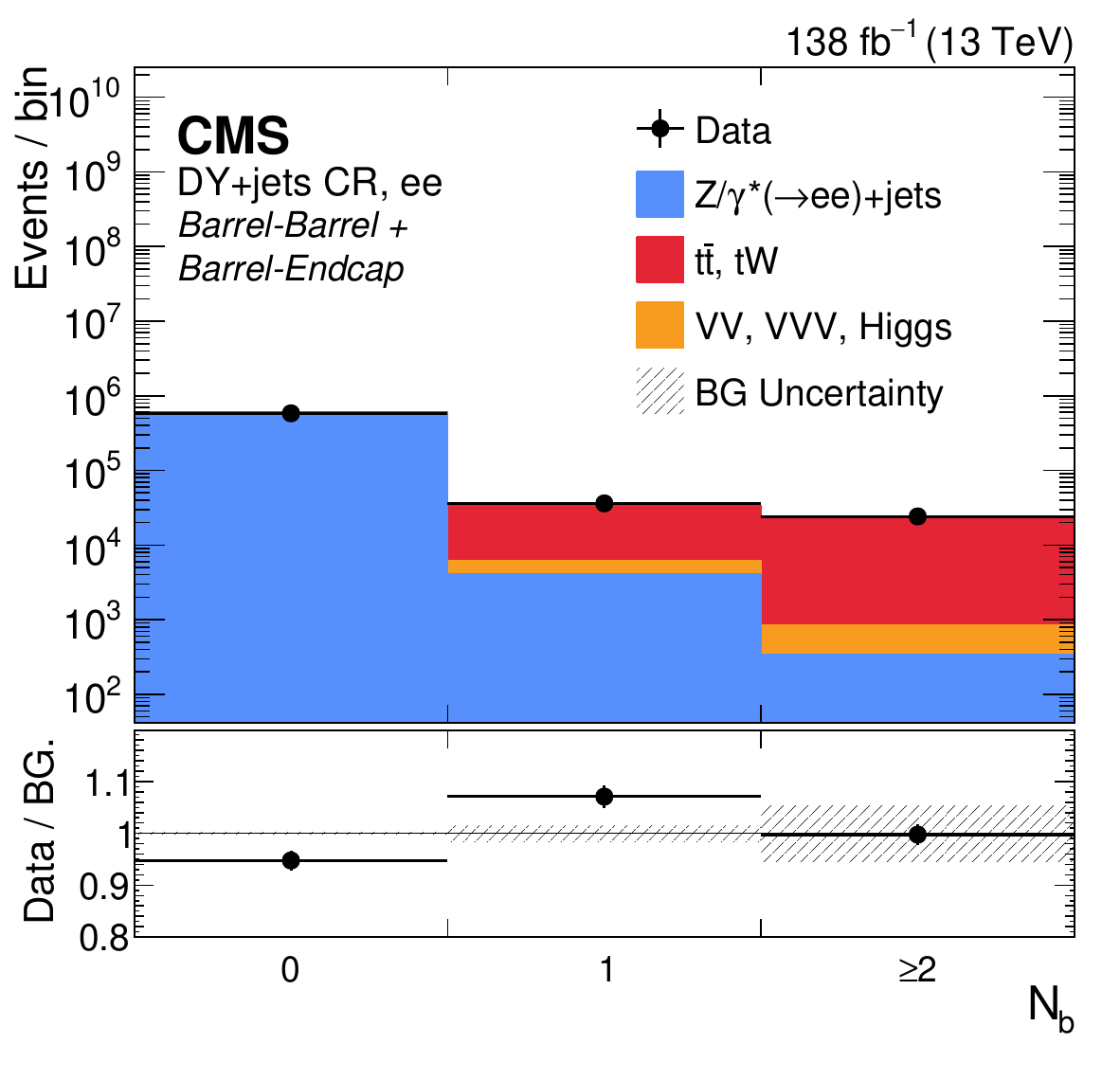}
  \includegraphics[width=0.45\textwidth]{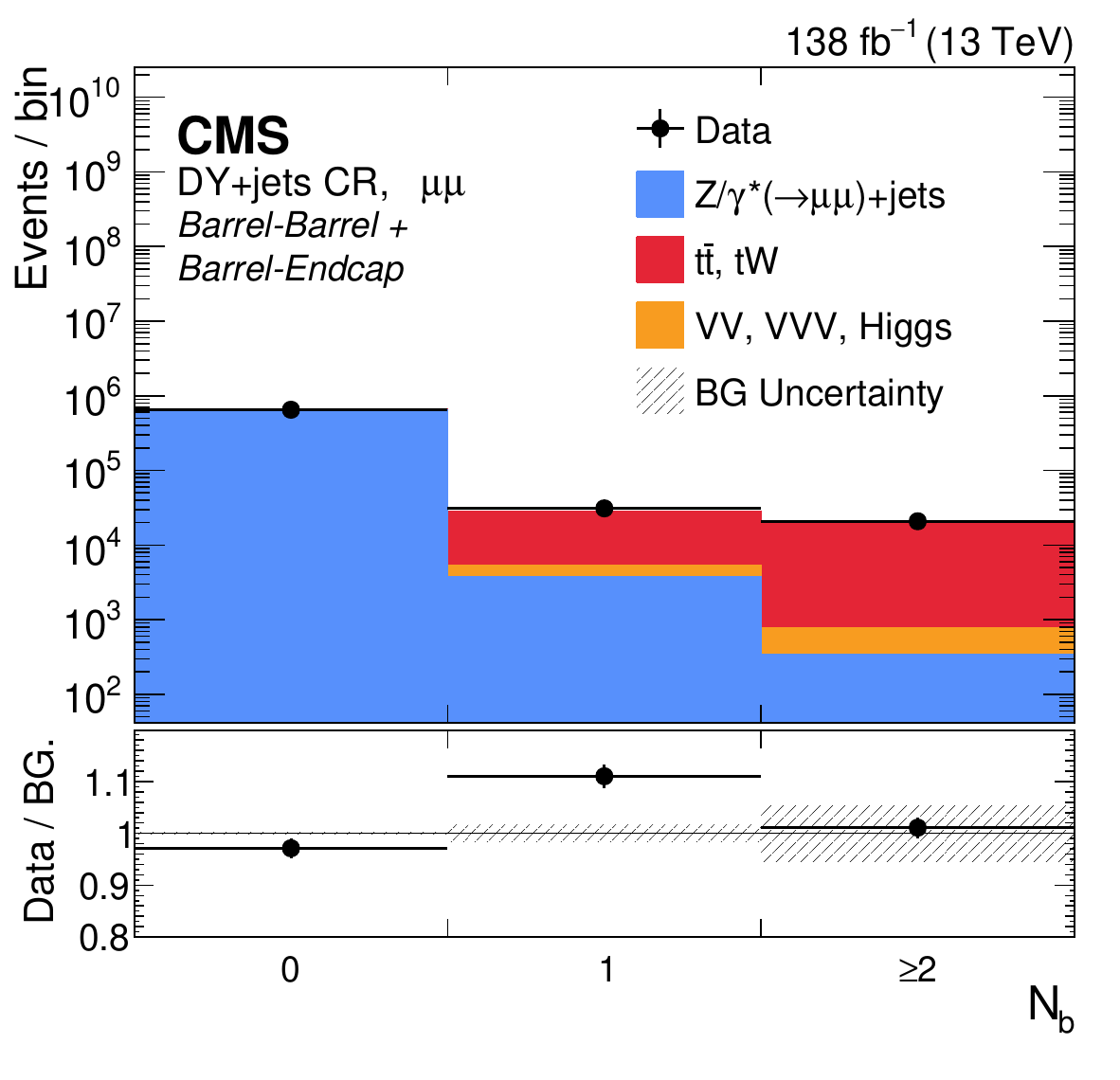}
  \caption{Comparison between data and various SM processes in DY+jets CR as a function of the number of \PQb jets in the $\Pe\Pe$ (left) and $\PGm\PGm$ (right) channels. The lower panel of each plot shows the ratio of data to predicted background yields. The hatched band in the lower panels indicates both the statistical and systematic uncertainty in the simulated background prediction.}
  \label{fig:DY_CR}
\end{figure}

After correcting the \ttbar MC sample, the residual differences observed in the DY+jets CR represent the mismodeling of the DY+jets MC sample. Similar to the \ttbar CR, the $\text{SF}_{\text{DY}}$ from the DY+jets CR are obtained as a function of \Nb. These $\text{SF}_{\text{DY}}$s are measured as a difference between data and all SM processes in the DY+jets CR. Figure~\ref{fig:DY_CR} shows the comparison between data and all SM processes in the DY+jets CR as a function of \Nb in the $\Pe\Pe$ (left) and $\PGm\PGm$ (right) channels. The data-to-simulation $\text{SF}_{\text{DY}}$s are derived separately for BB and BE categories for each data-taking year. In both the $\Pe\Pe$ and $\PGm\PGm$ channels, the $\text{SF}_{\text{DY}}$s show a slight dependence on \Nb, with values generally close to unity, indicating reasonable agreement between data and simulation. The $\text{SF}_{\text{DY}}$s in the 0\PQb final state range from 0.92 to 0.96, while slightly higher variations are observed in the 1\PQb and 2\PQb final states, where the $\text{SF}_{\text{DY}}$s range from 0.92 to 1.10. The $\text{SF}_{\text{DY}}$s in the $\Pe\Pe$ and $\PGm\PGm$ channels are consistent, showing no significant discrepancies between the two CRs. All the simulated backgrounds are corrected with the observed $\text{SF}_{\text{DY}}$, and the associated statistical uncertainty in the $\text{SF}_{\text{DY}}$ is taken to be the systematic uncertainty. An additional 10\% uncertainty in the $\text{SF}_{\text{DY}}$ is included in the systematic uncertainty for DY+jets background in the 1\PQb and 2\PQb final states to cover the deviations of data from prediction, and to take into account any bias because of the low number of events for the DY+jets process in these categories.

Background sources with nonprompt leptons and jets misidentified as leptons make very small contributions (1--3\%), to both the $\Pe\Pe$ and $\PGm\PGm$ channels. These backgrounds, jointly represented as ``Jets'' background in the following, are estimated using data CRs enriched in multijet events~\cite{CMS:2014lcz} by loosening the lepton identification requirements, in bins of $\pt$ and $\eta$ separately for the barrel and endcap regions. The fake rate is then applied to events in the SR where one or both leptons fail the tight identification but pass the preselection, after subtracting other background contributions estimated from simulation and correcting for double counting in dijet events. The method is validated in same-sign dilepton CRs for both the $\Pe\Pe$ and $\PGm\PGm$ channels. To account for possible mismodeling of fake contributions, a conservative 50\% systematic uncertainty is assigned to the ``Jets'' background estimate. Even though the uncertainty in these estimates is high, these backgrounds have a negligible effect on the results owing to their small contribution. 

\section{Systematic uncertainties}

The main experimental systematic uncertainties considered in this analysis include the uncertainty in the integrated luminosity, the modeling of pileup interactions, the lepton reconstruction, identification, and trigger efficiencies, as well as the energy scale and resolution of jets and the \PQb tagging efficiencies. Additional uncertainties affecting the background estimates, such as scale factors derived in CRs, are also included.

The systematic uncertainty in the integrated luminosity measurement is 1.2\% for 2016 data, 2.3\% for 2017 data, and 2.5\% for 2018 data, with an uncertainty of 1.6\% for the combined sum of the three years~\cite{CMS-LUM-17-003, CMS-PAS-LUM-17-004, CMS-PAS-LUM-18-002}. Since the dominant background processes are normalized in data CRs, this uncertainty mainly affects the signal simulation. Systematic uncertainties related to the pileup reweighting were assessed by recalculating the weights, with the assumed value of the total inelastic cross section shifted up and down by 4.6\% with respect to the nominal value~\cite{Sirunyan:2018nqx}. The effect of this uncertainty is very small on signal and background ($<$1\%).

The effect of the JES uncertainty on the estimated event yields is $\approx$2\% for signal and background in the 1\PQb final state and is around 2--4\% in the 2\PQb final state, whereas the effect of jet energy resolution (JER) uncertainty is negligible compared to the statistical uncertainty of the simulation. Uncertainties in the \PQb tagging efficiency are estimated by varying the associated SFs by one standard deviation (SD) and computing the impact on the estimated event yields for background and signal. The SF depends on the flavor of the parton that originated the jet and the uncertainties are estimated for heavy-flavor (\PQb and \PQc) jets and light-flavor quark or gluon jets separately. The effect of this uncertainty is less than 5\% for both signal and background. 

Systematic effects due to imprecise knowledge of the PDFs in the SR are determined by reweighting the simulated mass distributions according to the replicas of the NNPDF3.1 next-to-NLO PDF set and computing the 68\% interval of the observed changes, using the PDF4LHC prescription~\cite{Butterworth:2015oua}. In addition to the PDF variations, uncertainties from variations of the renormalization and factorization scales are also evaluated by varying the scales by factors of 0.5 and 2.0, respectively. The contribution of the PDF and scale uncertainties at low mass ($<$1500\GeV) is $\approx$2\% and at high mass is $\approx$5\% for the DY+jets simulation. The uncertainty is significant for the subdominant \ttbar background, since it is initiated predominantly by gluon fusion, resulting in an uncertainty of around 30\% at high mass. The combined PDF and scale uncertainty for signal samples is around 2\% at high masses.

The overall event selection efficiency including trigger, reconstruction, identification, and isolation is assigned a relative uncertainty of 6 (8)\% for barrel-barrel (barrel-endcap) events in the $\Pe\Pe$ channel, whereas an uncertainty of 1--2\% is assigned everywhere in the $\PGm\PGm$ channel. 

The uncertainty in the electron energy scale is 2\% in the barrel and 1\% for the endcap region for all three years of data taking. The uncertainty in the $\Pe\Pe$ mass resolution is negligible.

Potential biases in the muon momentum scale, particularly at high \pt, have been studied using the generalized endpoint method~\cite{highpt_muons}. No bias has been found within the statistical uncertainties in the method. These uncertainties are hence assigned as systematic uncertainties in the muon \pt scale. The impact of this uncertainty is 1--2\% for background events at high mass (3000--4000\GeV); whereas it is as large as 5\% for signal events. 

The uncertainty in the trigger prefiring rate in the forward region of the ECAL endcap in 2016 and 2017 is 20\%, resulting in a 2--3\% uncertainty in the acceptance for $\Pe\Pe$ events. The effect of trigger prefiring rate on the acceptance of $\PGm\PGm$ events is less than 1\%. 

The statistical uncertainties associated with the SFs measured in DY+jets and \ttbar CRs are assigned as systematic uncertainties to the corresponding background estimates in the SR and are less than 1\% in all cases. Additionally, the systematic uncertainty in the \ttbar estimation is derived from the deviations observed in the \ttbar VR in data and simulation after applying the DNN selection. Moreover, the systematic uncertainty based on shape differences between the \ttbar VR and SR is also considered. These uncertainties are applied in bins of dilepton invariant mass and range from 10--40\%.

The systematic uncertainties in the highest two mass bins of dilepton mass distribution for DY+jets background are shown in Table~\ref{tab:sys_elec_0b} (\ref{tab:sys_muon_0b}) for the $\Pe\Pe$ ($\PGm\PGm$) channels, respectively. 

\begin{table}[htp!]
  \centering
  \topcaption{Systematic uncertainties for the DY+jets background prediction in the highest two mass bins for the $\Pe\Pe$ channel. The values shown are post-fit, with all background scale factors applied. The uncertainties are evaluated inclusively across the \PQb jet multiplicity and $|\eta|$ regions. The dominant sources of uncertainty remain consistent across the different categories.}
  \begin{tabular}{l c c}
    &1100--1900\GeV & $>$1900\GeV \\ \hline
    DY+jets normalization  &1--10\% &1--10\% \\
    Electron trigger      &3.2\% &3.2\% \\
    Electron identification and isolation               &3.7\% &3.7\% \\
    Electron energy scale &8.4\% &11.5\% \\
    L1 prefiring          &$<$0.1\% &$<$0.1\% \\
    Pileup                 &0.4\% &0.2\% \\
    PDF                    &2.8\% &4.7\% \\
    JES             &2.0\% &2.0\% \\
    JER        &0.5\% &0.5\% \\
    \PQb tagging heavy-flavor SF  &3.0\% &4.0\% \\
    \PQb tagging light-flavor/gluon SF  &2.8\% &2.9\%
  \end{tabular}
  \label{tab:sys_elec_0b}
\end{table}

\begin{table}[htp!]
  \centering
  \topcaption{Systematic uncertainties for the DY+jets background prediction in the highest two mass bins for the $\PGm\PGm$ channel. The values shown are post-fit, with all background scale factors applied. The uncertainties are evaluated inclusively across the \PQb jet multiplicity and $|\eta|$ regions. The dominant sources of uncertainty remain consistent across the different categories.}
  \begin{tabular}{l c c}
    &1100--1900\GeV & $>$1900\GeV \\ \hline
    DY+jets normalization  &1--10\% &1--10\% \\ 
    Muon trigger  &1.6\% &2.0\% \\
    Muon reconstruction  &0.8\% &1.0\% \\
    Muon identification   &0.7\% &0.9\% \\
    Muon isolation  &0.7\% &0.9\% \\
    Muon mass scale  &4.8\% &5.5\% \\
    Muon resolution &0.1\% &0.4\% \\
    L1 prefiring  &0.1\% &0.1\% \\
    Pileup  &0.2\% &0.1\% \\
    PDF  &2.8\% &4.7\% \\
    JES  &2.0\% &2.0\% \\
    JER  &0.5\% &0.5\% \\
    \PQb tagging heavy-flavor SF  &3.0\% &4.0\% \\
    \PQb tagging light-flavor/gluon SF  &2.8\% &2.9\%
  \end{tabular}
  \label{tab:sys_muon_0b}
\end{table}

\section{Results}
\begin{figure}[tbh!]
  \centering
  \includegraphics[width=0.45\textwidth]{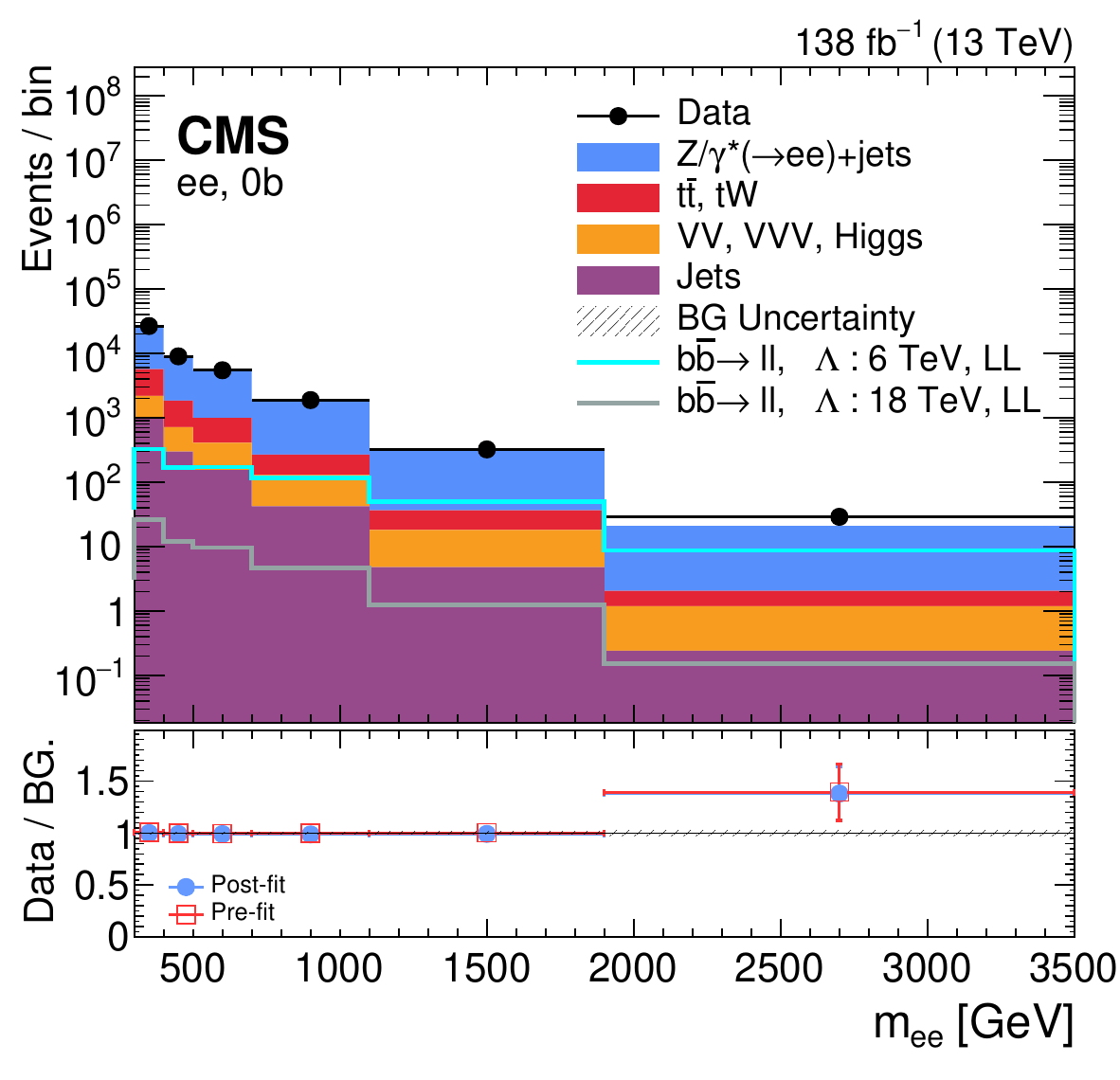}
  \includegraphics[width=0.45\textwidth]{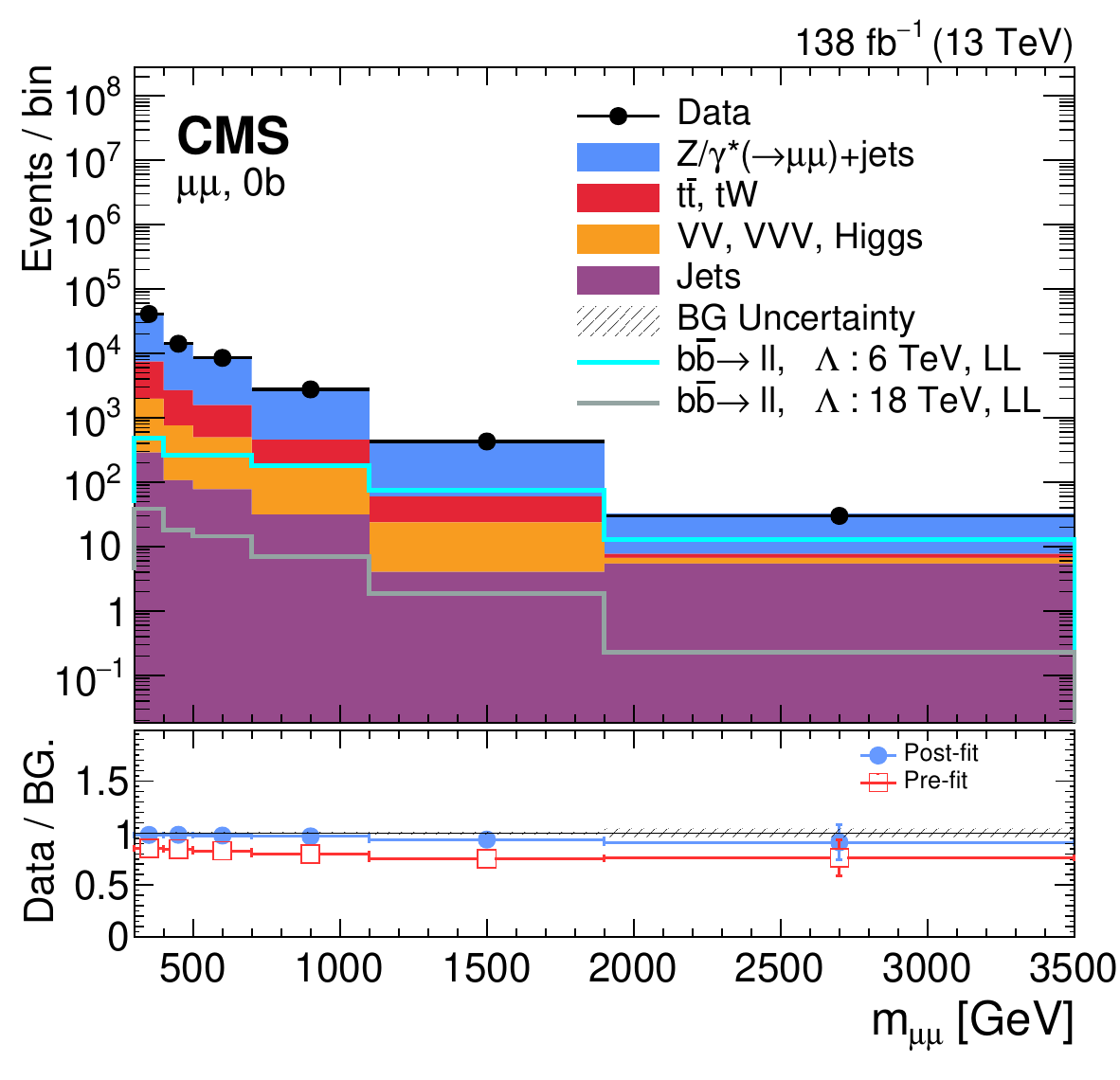}
  \caption{Observed \mll distributions in the data, and the post-fit backgrounds (stacked histograms), in the SR for $\Pe\Pe$+0\PQb (left) and $\PGm\PGm$+0\PQb (right) channels. The lower panels show ratios of the data to the pre-fit background prediction and post-fit background yield as red open squares and blue points, respectively. The hatched band in the lower panels indicates the systematic component of the post-fit uncertainty. The solid lines in the upper panels correspond to the \bbll signal expectations, for $\Lambda = 6$ and 18\TeV in the LL constructive interference model.}   
  \label{fig:total_bkgs1}
\end{figure}

\begin{figure}[tbh!]
  \centering
  \includegraphics[width=0.45\textwidth]{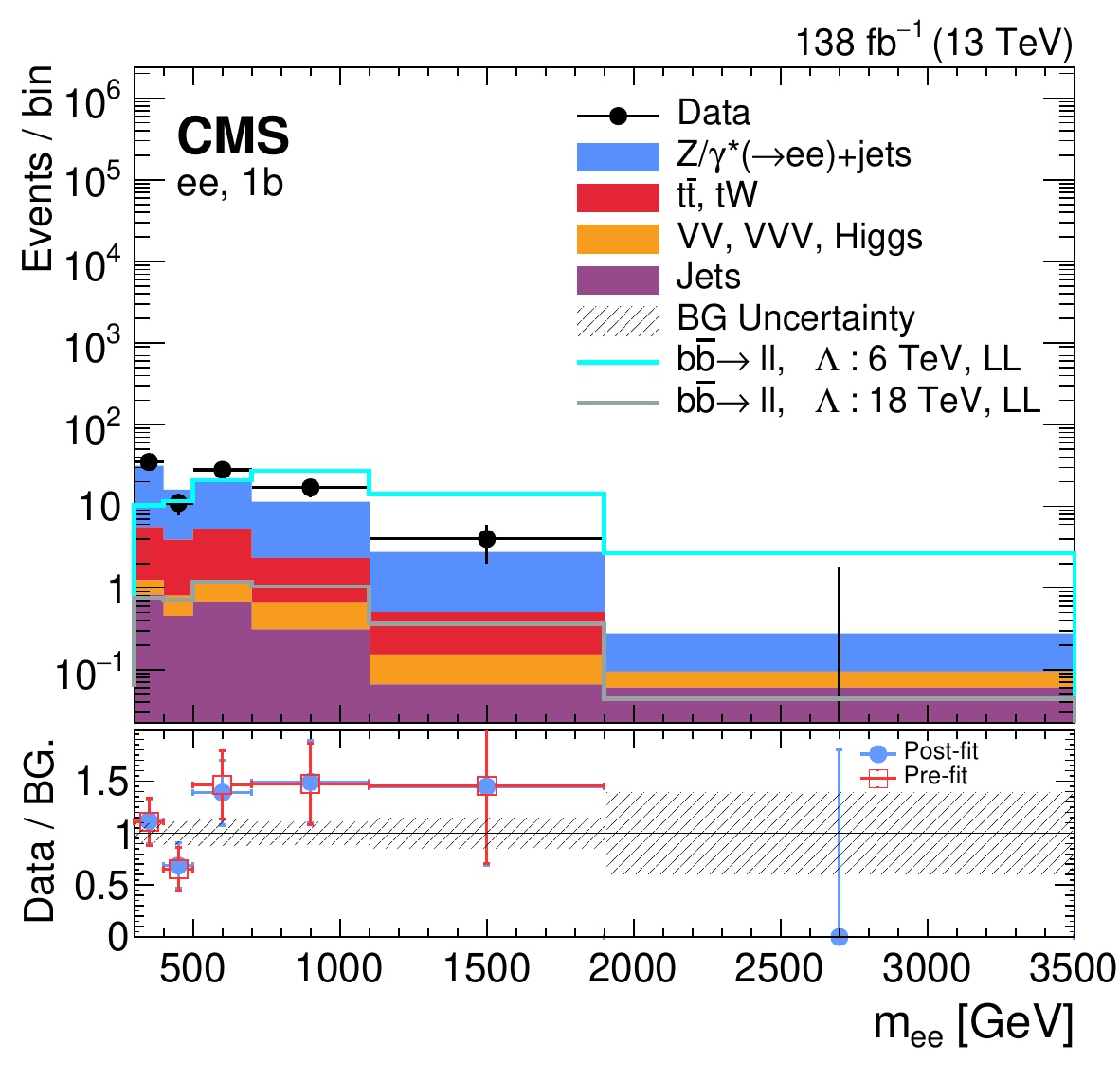} 
  \includegraphics[width=0.45\textwidth]{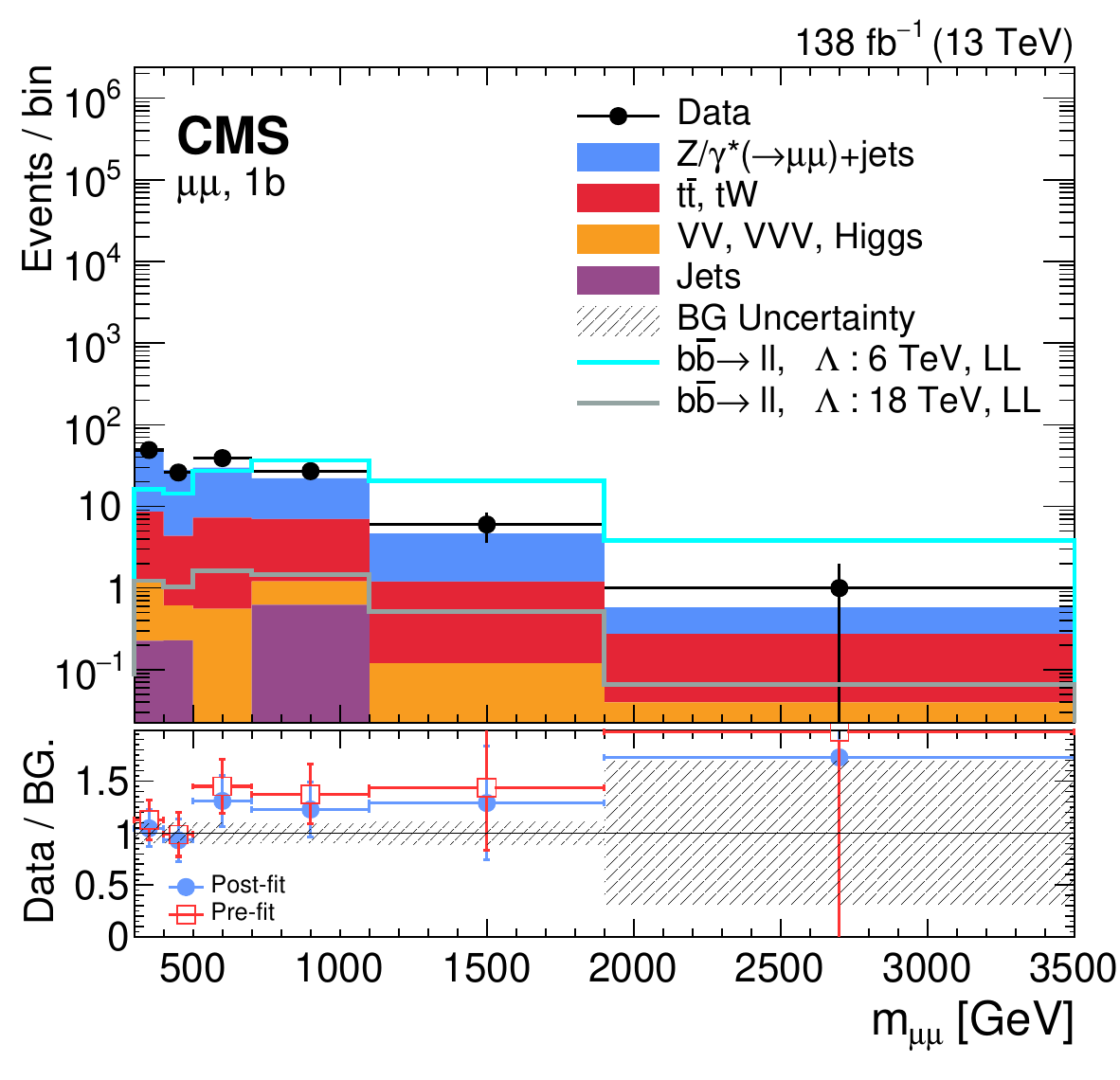} 
  \caption{Observed \mll distributions in the data, and the post-fit backgrounds (stacked histograms), in the SR for $\Pe\Pe$+1\PQb (left) and $\PGm\PGm$+1\PQb (right) channels. The lower panels show ratios of the data to the pre-fit background prediction and post-fit background yield as red open squares and blue points, respectively. The hatched band in the lower panels indicates the systematic component of the post-fit uncertainty. The solid lines in the upper panels correspond to the \bbll signal expectations, for $\Lambda = 6$ and 18\TeV in the LL constructive interference model.}   
  \label{fig:total_bkgs2}
\end{figure}

\begin{figure}[tbh!]
  \centering
  \includegraphics[width=0.45\textwidth]{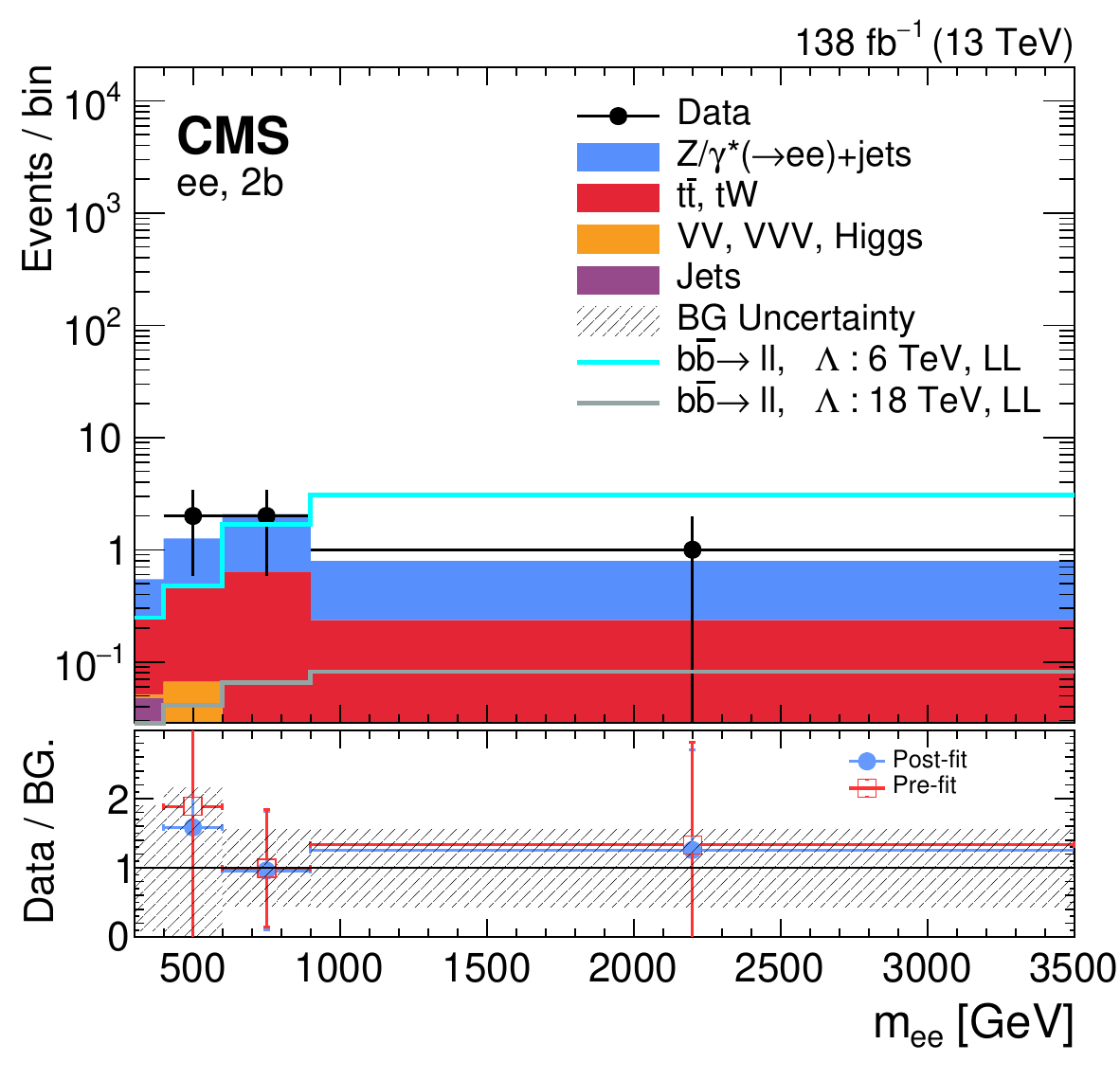} 
  \includegraphics[width=0.45\textwidth]{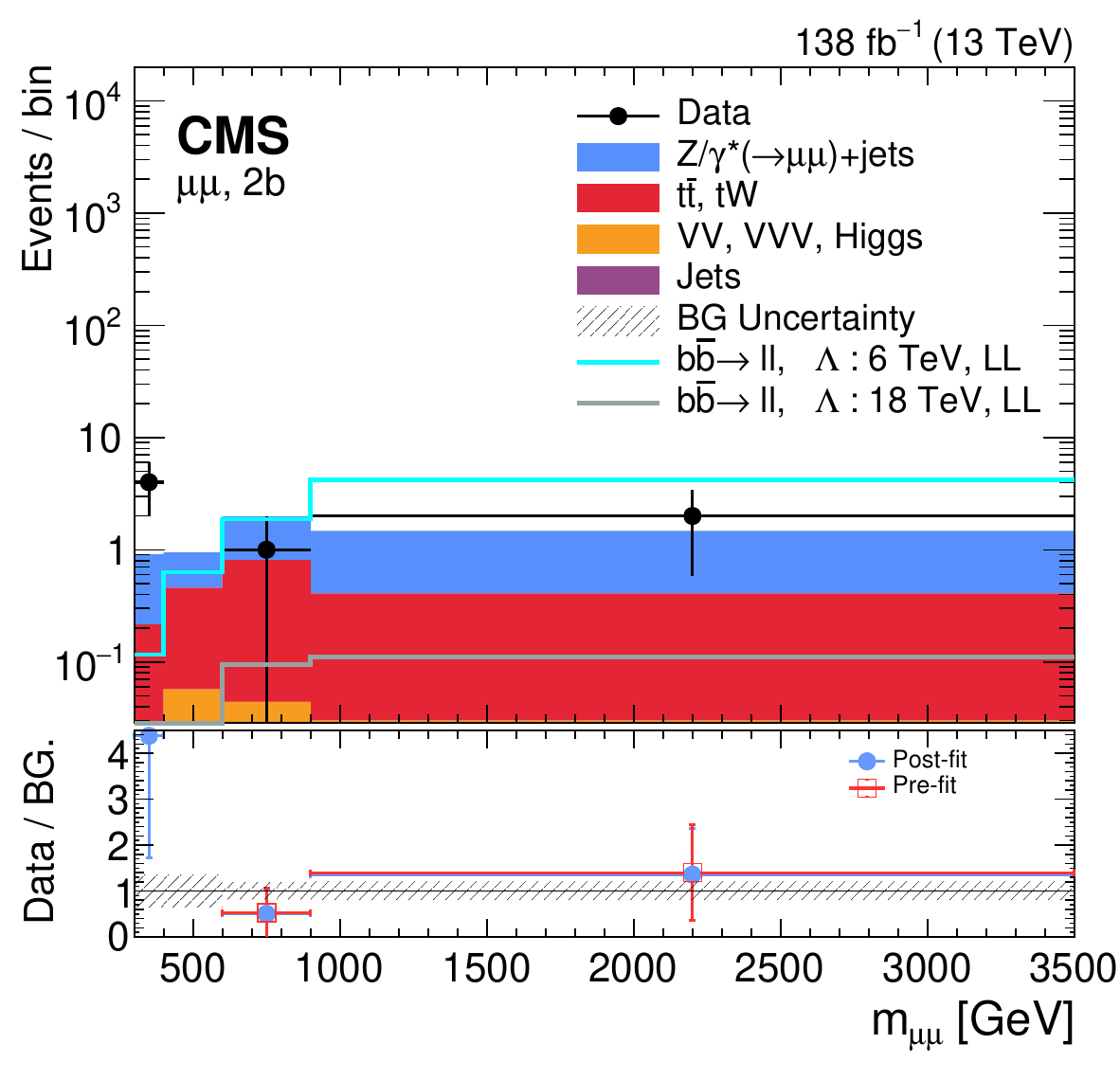} 
  \caption{Observed \mll distributions in the data, and the post-fit backgrounds (stacked histograms), in the SR for $\Pe\Pe$+2\PQb (left) and $\PGm\PGm$+2\PQb (right) channels. The lower panels show ratios of the data to the pre-fit background prediction and post-fit background yield as red open squares and blue points, respectively. The hatched band in the lower panels indicates the systematic component of the post-fit uncertainty. The solid lines in the upper panel correspond to the \bbll signal expectations, for $\Lambda = 6$ and 18\TeV in the LL constructive interference model.}   
  \label{fig:total_bkgs3}
\end{figure}

The invariant mass distributions in the $\Pe\Pe$ and $\PGm\PGm$ channels in the 0\PQb, 1\PQb, and 2\PQb final states are shown in Figs.~\ref{fig:total_bkgs1}--\ref{fig:total_bkgs3}, respectively. For illustration, the invariant mass distributions are overlaid for \bbll signals with $\Lambda = 6$ and 18\TeV, with the LL chirality configuration and constructive interference. To obtain the background predictions shown in these figures, a simultaneous background-only maximum-likelihood fit is performed across all SRs, separated by \PQb-tagged jet multiplicity and lepton $\eta$ categories. The parameter of interest is the background normalization, with signal strength fixed to zero. The observed mass spectra shown in the plots are consistent with the SM predictions. A slight excess, with a significance of less than 2 SD, is seen in the $\Pe\Pe$ channel for masses above 1.9\TeV (Fig.~\ref{fig:total_bkgs1} left). This is compatible with the previous measurement, without a \PQb-tagged jet requirement, performed with the Run~2 data set by the CMS experiment~\cite{CMS:2021ctt}, which showed a slight excess of events in the $\Pe\Pe$ channel for masses above 1.8\TeV.

\clearpage

Tables~\ref{tab:eeres1}--\ref{tab:eeres3} show the number of events for various mass ranges for the $\Pe\Pe$ and $\PGm\PGm$ channels in the 0\PQb, 1\PQb, and 2\PQb final states. The uncertainty in the total background is calculated taking into account correlations in the uncertainties among the different background sources. All values are post-fit and correspond to the SR.

\begin{table}[htp!]
  \centering
  \topcaption{Post-fit event yields for $\Pe\Pe$ and $\PGm\PGm$ channels in the 0\PQb final state, in the SR. The uncertainties include both statistical and systematic contributions.}
  \begin{tabular}{lcccccc}
    & \multicolumn{2}{c}{$\Pe\Pe$ channel}  & \multicolumn{2}{c}{$\PGm\PGm$ channel}       \\ 
    \mll [{\GeVns}] & Observed yield  & Total background  & Observed yield  & Total background             \\ \hline
    200--300            &     108\,042      & 107\,890 $\pm$ 940\,\,\,     & 161\,236          & 161\,000 $\pm$ 440               \\
    300--400            &     26\,639       & 26\,640 $\pm$ 240        & 40\,982           &\,\,\,41\,700 $\pm$ 120               \\
    400--500            &     9\,004        & 9\,093 $\pm$ 84          & 14\,135           & 14\,370  $\pm$ 53    \\
    500--700            &     5\,482        & 5\,536 $\pm$ 53          & 8\,530            & \,\,\,8\,738 $\pm$ 39  \\
    700--1100           &     1\,885        & 1\,910 $\pm$ 21          & 2\,762            & \,\,\,2\,852 $\pm$ 16  \\
    1100--1900          &     323         & \,325.6 $\pm$ 4.4        & 429             & \,\,\,\,457.6  $\pm$ 4.7  \\
    $>$1900           &     29          & \,\,\,\,21.01 $\pm$ 0.55     & 30              & \,\,\,\,\,\,\,33.0 $\pm$ 1.4
  \end{tabular}
  \label{tab:eeres1}
\end{table}

\begin{table}[htp!]
  \centering
  \topcaption{Post-fit event yields for $\Pe\Pe$ and $\PGm\PGm$ channels in the 1\PQb final state, in the SR. The uncertainties include both statistical and systematic contributions.}
  \begin{tabular}{lcccccc}
    & \multicolumn{2}{c}{$\Pe\Pe$ channel}  & \multicolumn{2}{c}{$\PGm\PGm$ channel}       \\ 
    \mll [{\GeVns}]  & Observed yield  & Total background       & Observed yield    & Total background        \\ \hline
    200--300            &     44          & 52.8 $\pm$ 4.3         &     127           & 116 $\pm$ 6.3\,\,\, \\
    300--400            &     35          & 31.4 $\pm$ 2.9         &     49            & 47 $\pm$ 4.2    \\
    400--500            &     11          & 16.0 $\pm$ 1.7         &     26            & 28 $\pm$ 2.7 \\
    500--700            &     28          & 20.2 $\pm$ 2.5         &     39            & 30 $\pm$ 2.9 \\
    700--1100           &     17          & 11.4 $\pm$ 1.2         &     27            & 22 $\pm$ 2.0 \\
    1100--1900          &     4           & \,\,\,2.76 $\pm$ 0.41      &     6             & 4.65 $\pm$ 0.51\,\, \\
    $>$1900             &     0           & \,\,\,0.28 $\pm$ 0.11      &     1             & 0.58 $\pm$ 0.40\,\,
  \end{tabular}
  \label{tab:eeres2}
\end{table}

\begin{table}[htp!]
  \centering
  \topcaption{Post-fit event yields for $\Pe\Pe$ and $\PGm\PGm$ channels in the 2\PQb final state, in the SR. The uncertainties include both statistical and systematic contributions.}
  \begin{tabular}{lcccccc}
    & \multicolumn{2}{c}{$\Pe\Pe$ channel}  & \multicolumn{2}{c}{$\PGm\PGm$ channel}       \\ 
    \mll [{\GeVns}] & Observed yield  & Total background    & Observed yield  & Total background           \\ \hline
    200--400            &     2           & 2.03 $\pm$ 0.95        &     4           & 2.53 $\pm$ 0.68            \\
    400--600            &     2           & 1.3 $\pm$ 1.2          &     0           & 0.94 $\pm$ 0.31            \\
    600--900            &     2           & 2.20 $\pm$ 0.95        &     1           & 1.90 $\pm$ 0.27            \\
    $>$900              &     1           & 0.81 $\pm$ 0.35        &     2             & 1.46 $\pm$ 0.08
  \end{tabular}
  \label{tab:eeres3}
\end{table}

The analysis also tests for lepton flavor universality by measuring the ratio of the differential production cross sections in the $\PGm\PGm$ and $\Pe\Pe$ channels (\Rmumuee) as a function of the dilepton invariant mass. The $\Pe\Pe$ and $\PGm\PGm$ mass distributions are not directly comparable because of the differences in the detector acceptance and lepton efficiencies, as well as the differences in mass scale and resolution between the final states. To address these limitations, the measurement is performed by first correcting the bin-by-bin migration effects due to mass scale and resolution in the $\Pe\Pe$ and $\PGm\PGm$ reconstruction-level mass distributions using the \textsc{TUnfold} package~\cite{Schmitt:2012kp} and then correcting the resulting corrected distribution for detector acceptance and lepton efficiency differences.

The bin-by-bin migration effects are corrected from the reconstruction-level distribution to generator level by using a 2-dimensional histogram of reconstructed versus generated mass (response matrix). The response matrices are obtained from the DY+jets MC events passing the full SR selections for each \PQb-tagged multiplicity region, eta category, data-taking year, and lepton flavor. For the lepton flavor ratio study, the 1\PQb and 2\PQb final states are merged, as the event yields in the 2\PQb region are limited and combining the two improves the statistical precision of the measurement. The reconstructed mass distribution in data after subtracting all SM backgrounds except for DY+jets in each channel is corrected by applying the inverted response matrices. Since the off-diagonal elements of the response matrices are typically at the level of 5--10\% relative to the diagonal entries, no regularization is necessary. The resulting corrected mass distribution is compared with the generator-level mass spectra for validation and is in good agreement. 

To account for the differences in detector acceptance and lepton efficiencies between the electron and muon channels, a normalization factor is derived, assuming no signal is present in the 200--400\GeV mass range, by taking the ratio of the corrected $\PGm\PGm$ mass distribution to the corrected $\Pe\Pe$ mass distribution in data. Similarly, a normalization factor is derived for the DY+jets simulation using the ratio of $\PGm\PGm$ and $\Pe\Pe$ events in the corrected simulated DY+jets mass distribution after full SR selections. Since this normalization is dependent on the jet multiplicity, we derive it for 0\PQb and $\geq$1\PQb jet final states separately, and independently for each data-taking year and $\eta$ category. The uncertainty in the factor is taken from the statistical uncertainty of a degree-0 polynomial fit in the 200--400\GeV range and propagated as an additional uncertainty in the ratio. The observed flavor ratio from the corrected mass distributions in data and DY+jets simulation is then corrected with their corresponding normalization factors to obtain the final flavor ratio from data and the DY+jets simulation. As detector acceptance and lepton efficiency differences increase with dilepton invariant mass, the flavor ratio for DY+jets simulation deviates from unity at higher masses as expected.

Most of the systematic uncertainties such as the choice of PDF set, background-specific CR SFs, JES, JER, and \PQb tagging scale factors cancel in the ratio, whereas the remaining systematic uncertainties affecting the mass due to modeling of the mass resolution and scale are propagated to the bin-by-bin migration correction step by obtaining additional response matrices with reconstructed mass values shifted by the uncertainty. The uncertainties in the detector acceptance and lepton efficiencies are propagated to the final flavor ratio via the acceptance and efficiency corrections. In addition, the uncertainties in the normalization factor are also propagated to the final flavor ratio. The dominant systematic contributions across all final states arise from the modeling of the lepton energy scale and resolution, while in the $\geq$1\PQb jet final state, the limited number of data events provides the largest overall source of uncertainty. 

\begin{figure}[htb!]
  \centering
  \includegraphics[width=0.45\textwidth]{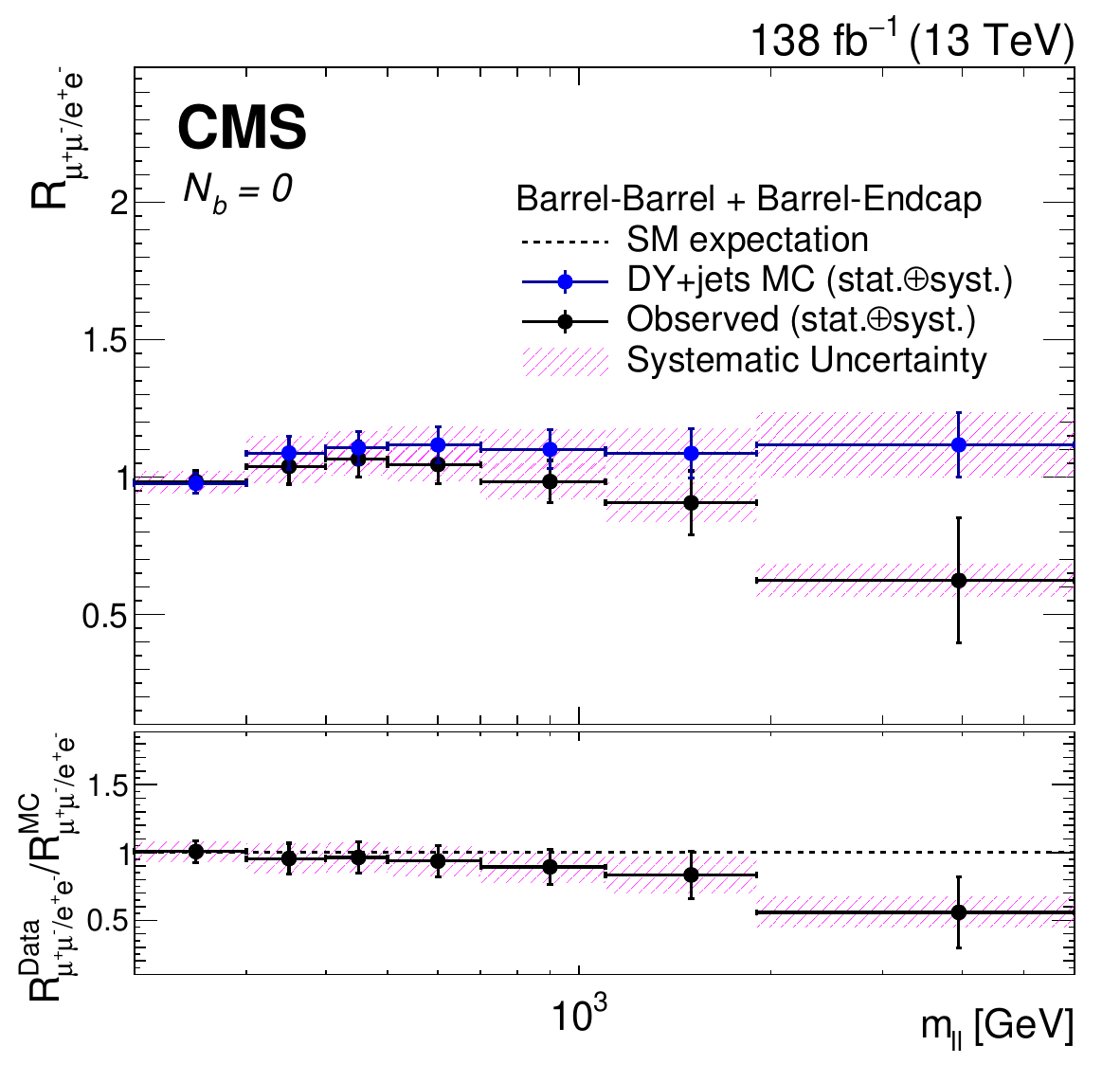}
  \includegraphics[width=0.45\textwidth]{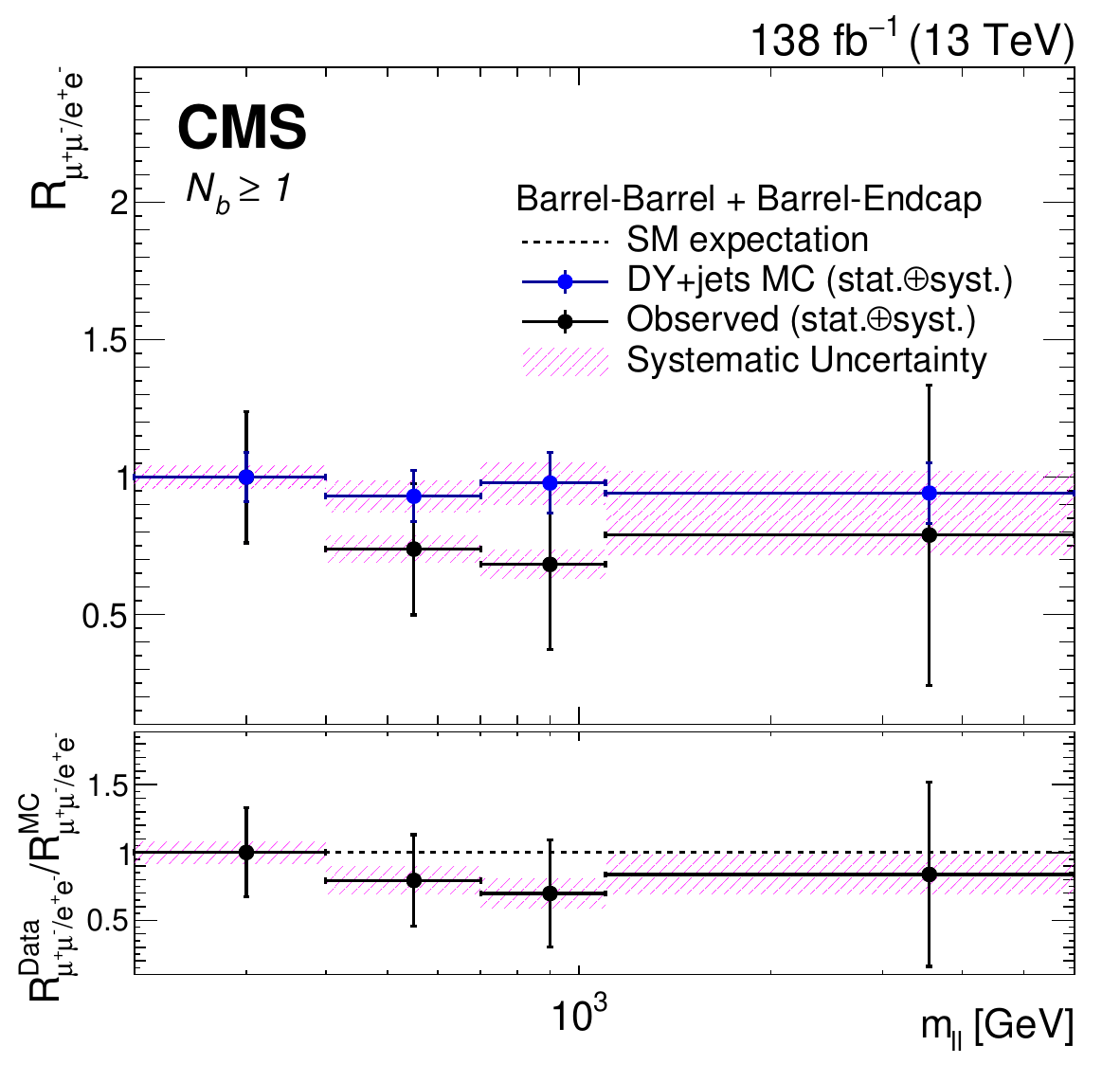}
  \caption{Ratio of the differential dilepton production cross section in the $\PGm\PGm$ and $\Pe\Pe$ channels as a function of dilepton mass in the 0\PQb (left) and $\geq$1\PQb (right) final states. The ratio is obtained after correcting the $\PGm\PGm$ and $\Pe\Pe$ reconstructed mass spectra for bin-by-bin migration effects due to mass scale and resolution. The upper panel shows the observed flavor ratio in data and the DY+jets simulation. The lower panel shows the flavor ratio observed in data and in DY+jets simulation after correcting for the residual differences in acceptance and efficiency as a function of mass. The error bars include both statistical and systematic uncertainties. The shaded band represents the systematic uncertainties in the estimation.}
  \label{fig:lep_ratio_0b}
\end{figure}

The observed lepton flavor ratio in the 0\PQb final state is shown in Fig.~\ref{fig:lep_ratio_0b} (left). The upper panel shows the observed flavor ratio in data and in DY+jets simulation. The deviation of the DY+jets simulation flavor ratio from the SM expectation is due to the difference in the acceptance and efficiency of $\Pe\Pe$ and $\PGm\PGm$ events as a function of mass. The observed flavor ratio is in good agreement with the DY+jets simulation. The small deviation visible at high mass is caused by the slight excess of $\Pe\Pe$ events noted in Fig.~\ref{fig:total_bkgs1}, which affects the highest bin of the measured flavor ratio. The lower panel shows the ratio of flavor ratios observed in data and in the DY+jets simulation to correct for the residual differences in acceptance and efficiencies as a function of mass. The observed lepton flavor ratio from corrected $\Pe\Pe$ and $\PGm\PGm$ mass spectra in the $\geq$1\PQb jet final state can be found in Fig.~\ref{fig:lep_ratio_0b} (right). Because of limited event counts in the 2\PQb final state, the ratio is estimated by combining the 1\PQb and 2\PQb final states. The observed ratio is consistent with that from the DY+jets simulation. A $\chi^2$ test for the mass range above 400\GeV is performed to quantify the agreement of the observed result with the measurement from the DY+jets simulation. The resulting $\chi^2$/dof value in the 0\PQb final state is $10.86/5$, and the one-sided $p$-value is 0.05. For the $\geq$1\PQb jet final state, the $\chi^2$/dof value is 0.99/3, giving the one-sided $p$-value as 0.80.

\section{Statistical interpretation}
Lower limits on the energy scale $\Lambda$ are obtained by comparing the upper limits on the signal cross section with the prediction calculated with \MGvATNLO at LO. The upper limits on the cross section are computed at 95\% confidence level (\CL) with Bayesian techniques using the CMS \textsc{Combine} tool~\cite{CMS:2024onh}. The limits are estimated separately for $\Pe\Pe$ and $\PGm\PGm$ channels, each time combining all data-taking years and $\abs{\eta}$ categories. The \PQb~jet categories are treated differently for the two signals. For the signal cross section a uniform prior is used, whereas the nuisance parameters are modeled with log-normal priors. A binned template based approach is used where the likelihoods are constructed in bins of \mll. For the models with destructive interference, the signal yield is obtained after subtracting the pure LO DY+jets contribution. The interference term becomes dominant for lower mass bins and higher $\Lambda$ values, therefore the limits are estimated using events above a certain dilepton invariant mass threshold. This mass threshold is optimized to obtain the best expected limits and is found to be 1000\GeV. A combination of the $\Pe\Pe$ and $\PGm\PGm$ channels is also performed, assuming identical branching fractions into electrons and muons for both signal models. 

\begin{figure}[tbh!]
  \centering
  \includegraphics[width=0.45\textwidth]{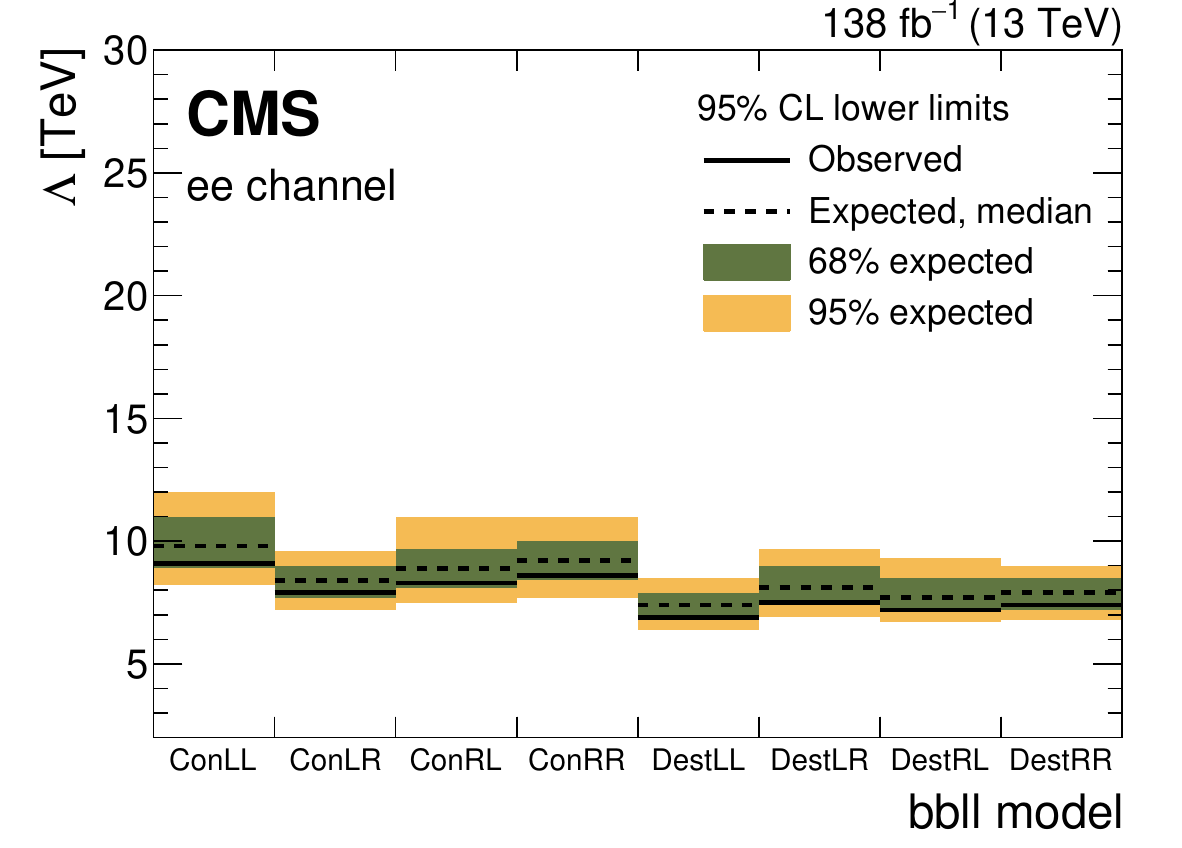}
  \includegraphics[width=0.45\textwidth]{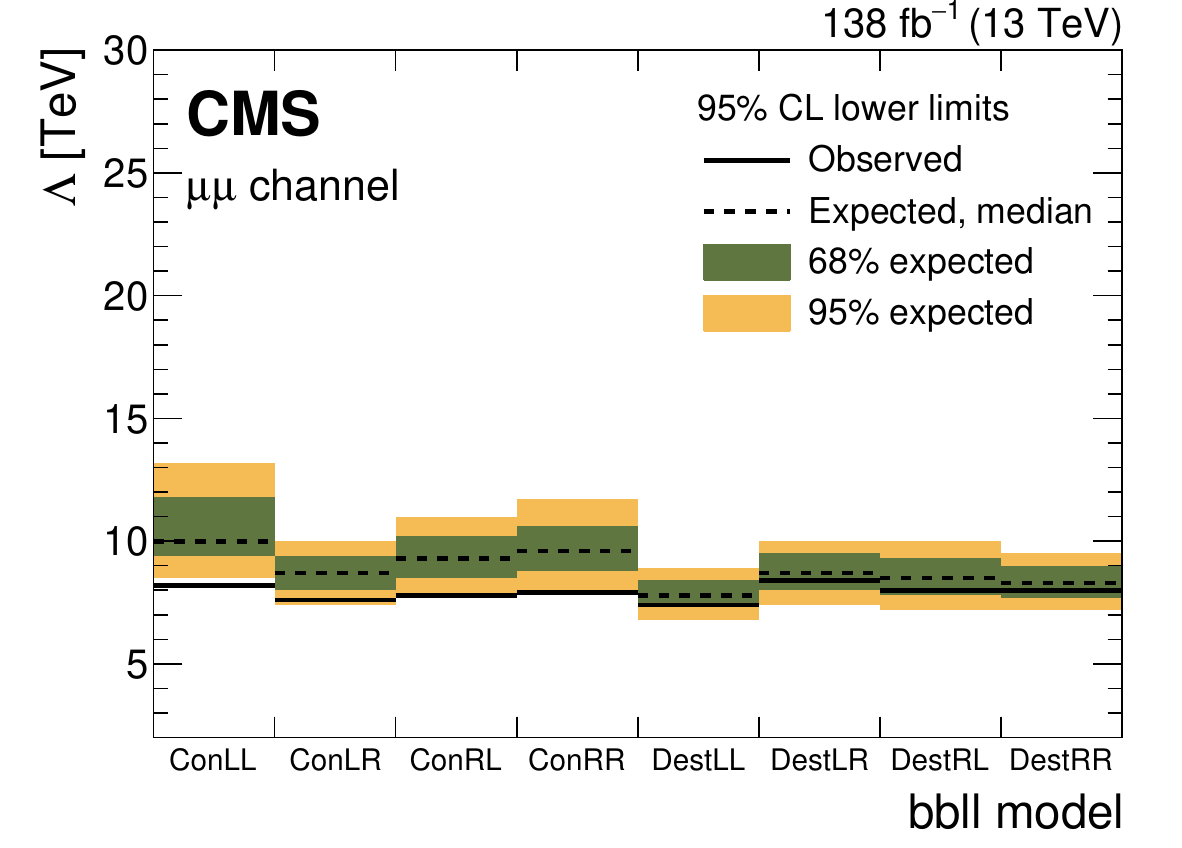}\\
  \includegraphics[width=0.45\textwidth]{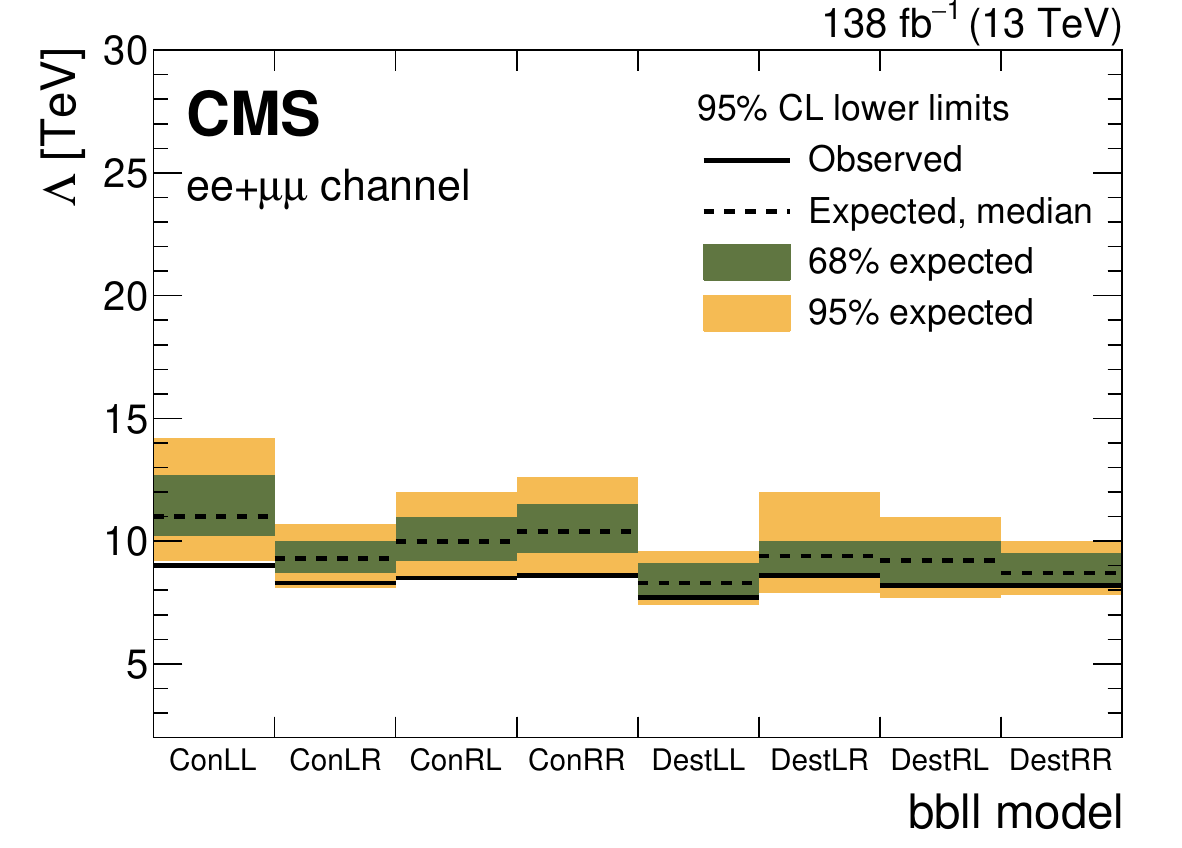}
  \caption{The 95\% \CL expected and observed lower limits on the energy scale $\Lambda$ for the \bbll model with different chirality and interference assumptions, namely constructive left-left (ConLL), left-right (ConLR), right-left (ConRL), right-right (ConRR), destructive left-left (DesLL), left-right (DesLR), right-left (DesRL), and right-right (DesRR) in the $\Pe\Pe$ (upper left), $\PGm\PGm$ (upper right) channels, and the combination (lower) with $\geq$0 \PQb-tagged jets. The shaded bands correspond to the 68\% and 95\% quantiles for the expected limits.}
  \label{fig:limits_bbll_ee}
\end{figure}

For the \bbll signal, the results for the 0\PQb, 1\PQb, and 2\PQb final states are combined in the limit calculation for a particular channel. Figure~\ref{fig:limits_bbll_ee} (upper left) shows the observed lower limits on $\Lambda$ for the \bbll signal in the $\Pe\Pe$ final state with constructive and destructive interference and different chirality assumptions. The exclusion for the different chirality and interference assumptions varies within 1 SD. The observed lower limits on $\Lambda$ for the same model in the $\Pgm\Pgm$ channel are shown in Fig.~\ref{fig:limits_bbll_ee} (upper right). Here, we can see that the exclusions vary within 2 SDs for different chirality and interference assumptions. The difference in observed and expected limits in the $\Pe\Pe$ and $\PGm\PGm$ channels is due to the slight excess in the 1\PQb final state mass spectra as shown in Fig.~\ref{fig:total_bkgs2}. Since the 1\PQb final state is the most sensitive channel, followed by the 0\PQb and 2\PQb channels, this leads to the lower observed limit in both the channels as shown in Fig.~\ref{fig:limits_bbll_ee}. The observed lower limits on $\Lambda$ for the same model with combined channels are shown in Fig.~\ref{fig:limits_bbll_ee} (lower plot). The expected and observed lower limits on $\Lambda$ in the \bbll signal are summarized in Table~\ref{tab:bblllimits}. The limits in the \bbll signal are stronger than the LEP ($\Pe^+\Pe^- \to \PQb\PAQb$) bounds~\cite{ALEPH:2006bhb, ALEPH:2013dgf} for the RR/LR/RL chiral structures and both constructive and destructive interference, while for the LL scenario the LEP constraints are tighter. 

\begin{table}[tbh!]
  \centering
  \topcaption{Lower limits at 95\% \CL on the energy scale $\Lambda$ in the \bbll signal model in \TeVns for constructive left-left (ConLL), left-right (ConLR), right-left (ConRL), right-right (ConRR), destructive left-left (DesLL), left-right (DesLR), right-left (DesRL), and right-right (DesRR) chirality and interference assumptions in the $\Pe\Pe$ and $\PGm\PGm$ channels and the combination of the two.}
  \cmsTable{
    \begin{tabular}{lccccccccc}
      &                                               & ConLL & ConLR    & ConRL  & ConRR  & DesLL & DesLR & DesRL & DesRR\\ \hline
      \multirow{ 2}{*}{$\Pe\Pe$}   &  Expected [{\TeVns}]    & 9.8   & 8.4      & 8.9    & 9.2    & 7.4   & 8.1   & 7.7   & 7.9 \\
      &  Observed [{\TeVns}]    & 9.1   & 7.9      & 8.3    & 8.6    & 6.9   & 7.5   & 7.2   & 7.4  \\
      \multirow{ 2}{*}{$\Pgm\Pgm$}   &  Expected [{\TeVns}]  & 10.0   & 8.7     & 9.3    & 9.6    & 7.8   & 8.7   & 8.5   & 8.3   \\
      &  Observed [{\TeVns}]    & 8.2   & 7.6      & 7.8    & 7.9    & 7.4   & 8.4   & 8.0   & 8.0 \\
      \multirow{ 2}{*}{Combined}   &  Expected [{\TeVns}]    & 11.0   & 9.3      & 10.0    & 10.4 & 8.3   & 9.4   & 9.2   & 8.7     \\
      &  Observed [{\TeVns}]    & 9.0   & 8.3      & 8.5    & 8.6    & 7.7   & 8.6   & 8.2   & 8.2
    \end{tabular}
  }
  \label{tab:bblllimits}
\end{table}

\begin{figure}[tbh!]
  \centering
  \includegraphics[width=0.45\textwidth]{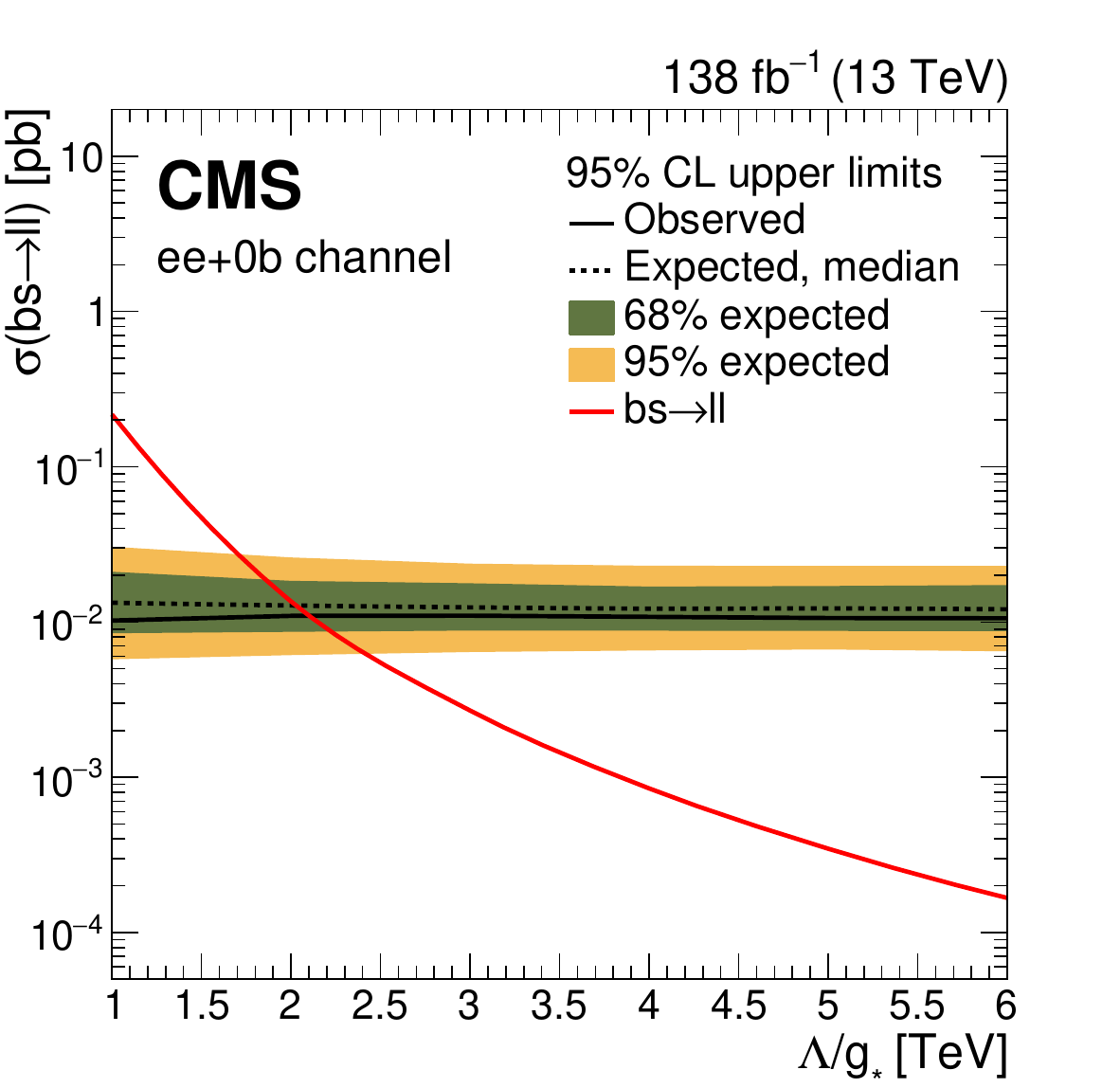}
  \includegraphics[width=0.45\textwidth]{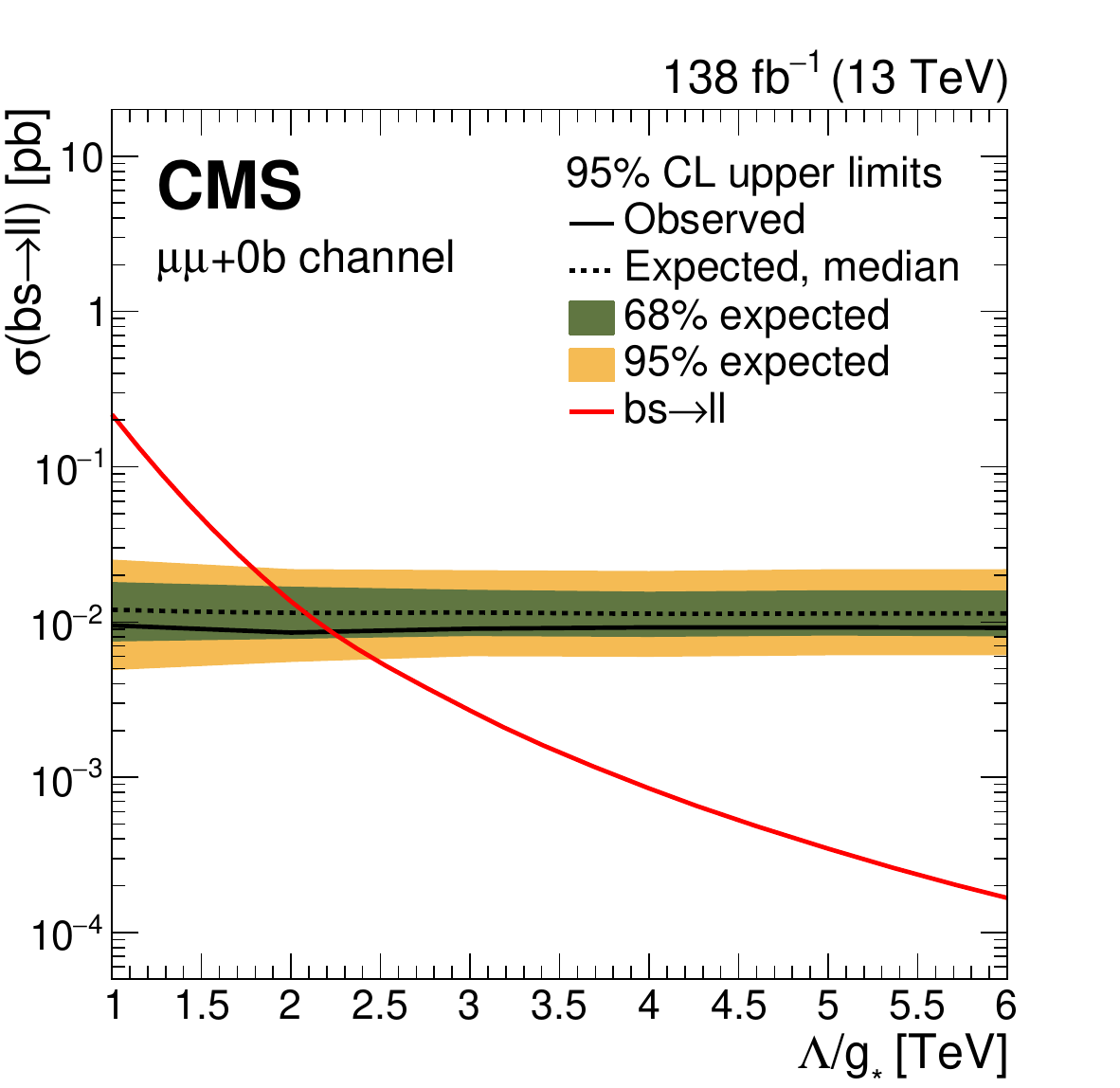}
  \caption{Upper limits at 95\% \CL on the production cross section for the \bsll signal in the $\Pe\Pe$ (left) and $\Pgm\Pgm$ (right) channels for 0 \PQb-tagged jets. The shaded bands correspond to the 68\% and 95\% quantiles for the expected limits. The red line corresponds to the theoretical prediction for the \bsll signal~\cite{Afik:2018nlr}.}
  \label{fig:bsll_limit_ind1}
\end{figure}

\begin{figure}[tbh!]
  \centering
  \includegraphics[width=0.45\textwidth]{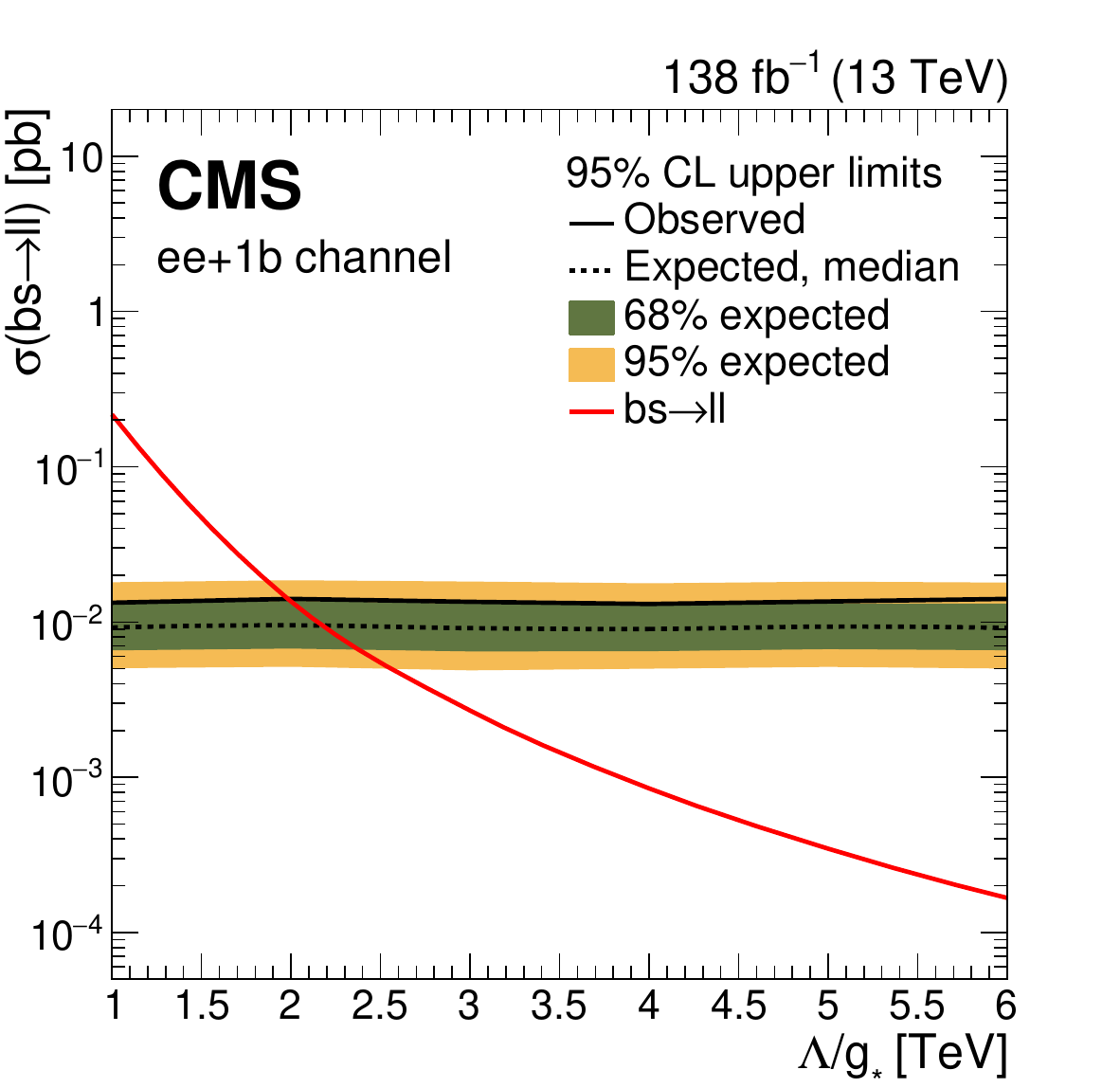}
  \includegraphics[width=0.45\textwidth]{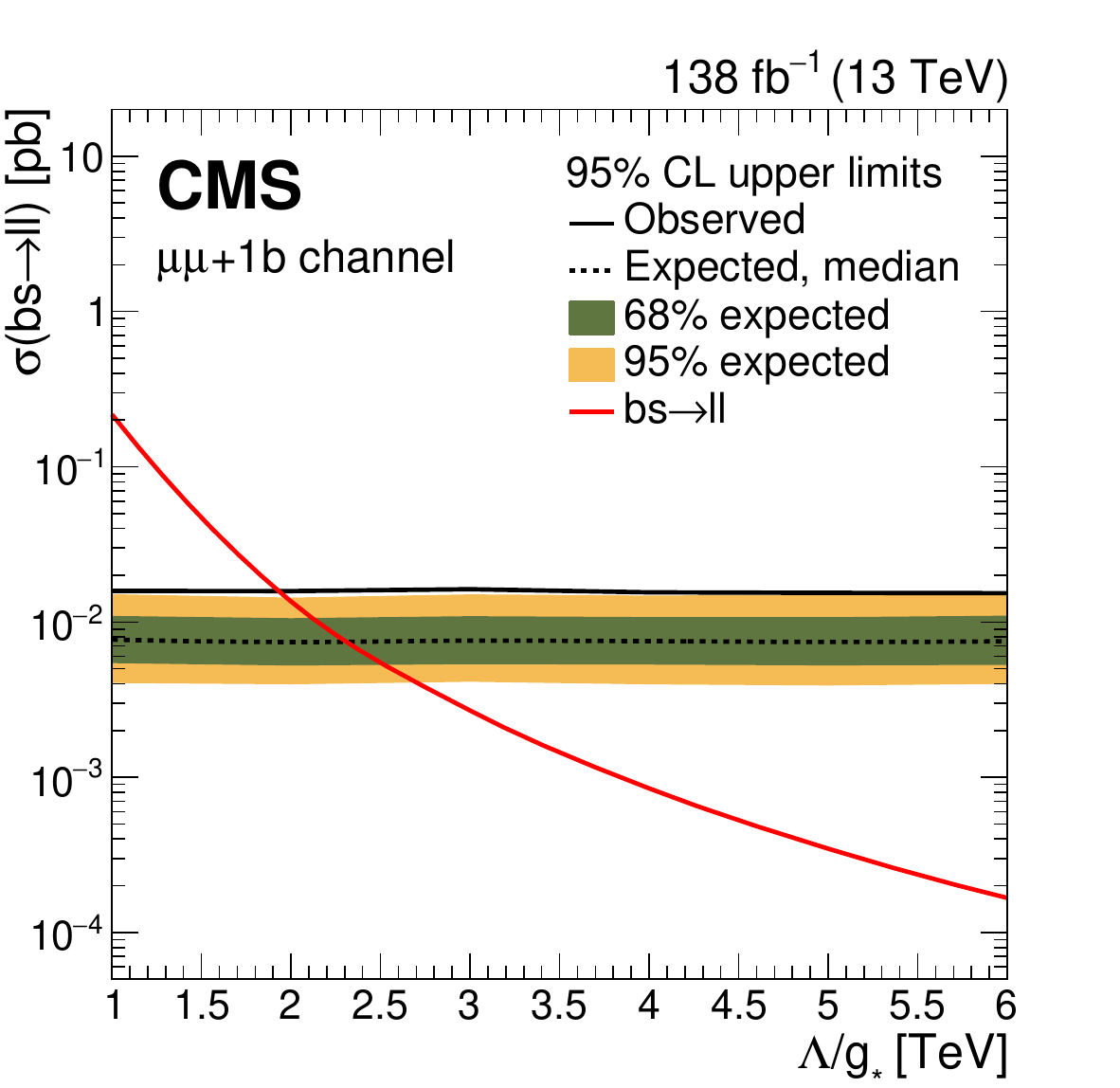}
  \caption{Upper limits at 95\% \CL on the production cross section for the \bsll signal in the $\Pe\Pe$ (left) and $\Pgm\Pgm$ (right) channels for 1 \PQb-tagged jets. The shaded bands correspond to the 68\% and 95\% quantiles for the expected limits. The red line corresponds to the theoretical prediction for the \bsll signal~\cite{Afik:2018nlr}.}
  \label{fig:bsll_limit_ind2}
\end{figure}

For the \bsll signal, the limit is calculated using the 0\PQb and 1\PQb final states. Figure~\ref{fig:bsll_limit_ind1} shows the observed limit as a function of \lamdagstar for the $\Pe\Pe$+0\PQb (left) and $\Pgm\Pgm$+0\PQb (right) channels. Figure~\ref{fig:bsll_limit_ind2} shows the observed limit as a function of \lamdagstar for the $\Pe\Pe$+1\PQb (left) and $\Pgm\Pgm$+1\PQb (right) channels. An observed lower limit on \lamdagstar of 2.1~(2.0)\TeV is found for the $\Pe\Pe$+0\PQb~(1\PQb) channel. For the $\PGm\PGm$+0\PQb~(1\PQb) channel, a lower limit of 2.2~(1.9)\TeV is observed on \lamdagstar. A 1 (2) SD nonresonant excess in the $\Pe\Pe$ ($\Pgm\Pgm$)+1\PQb channel is observed for the \bsll signal. This slight excess arises because of an increased number of data events, as shown in Fig.~\ref{fig:total_bkgs2}, which remain consistent with the predicted background within statistical and systematic uncertainties. The lower limits on \lamdagstar in this model for the 0\PQb, 1\PQb, and combined 0\PQb and 1\PQb channels are summarized in Table~\ref{tab:bslllimits1} for the $\Pe\Pe$, $\PGm\PGm$, and combined channels. A value of 2.4\TeV is found as the observed lower limit on \lamdagstar for the combination of the two channels. This result is comparable to the existing limits from the LHC experiments~\cite{ATLAS:2021mla}.

\begin{table}[tbh!]
  \centering
  \topcaption{Lower limits at 95\% \CL on \lamdagstar in the \bsll signal model in \TeVns for 0\PQb, 1\PQb, and combined (0\PQb+ 1\PQb) channels. Shown are the expected and observed limits in the $\Pe\Pe$ and $\PGm\PGm$ channels and the combination of the two.}
  \begin{tabular}{lccccc}
    &                                               & 0\PQb & 1\PQb    & combined (0\PQb+ 1\PQb)  \\ \hline
    \multirow{ 2}{*}{$\Pe\Pe$}   &  Expected [{\TeVns}]    & 2.0   & 2.2      & 2.3  \\
    &  Observed [{\TeVns}]    & 2.1   & 2.0      & 2.2  \\
    \multirow{ 2}{*}{$\Pgm\Pgm$}   &  Expected [{\TeVns}]  & 2.1   & 2.3      & 2.4  \\
    &  Observed [{\TeVns}]    & 2.2   & 1.9      & 2.2  \\
    \multirow{ 2}{*}{Combined}   &  Expected [{\TeVns}]    & 2.2   & 2.5      & 2.6  \\
    &  Observed [{\TeVns}]    & 2.4   & 2.0      & 2.4
  \end{tabular}

  \label{tab:bslllimits1}
\end{table}

\section{Summary}

A search for new physics in the high-mass dilepton final state produced in association with \PQb-tagged jets, focusing on nonresonant phenomena, has been performed using the proton-proton collision data collected during 2016--2018, corresponding to an integrated luminosity of 138\fbinv. Two models of nonresonant signatures have been considered. For a four-fermion contact interaction in \bbll ($\Pell\Pell$, electrons or muons) production, lower limits on the scale of new physics ($\Lambda$) are set depending on the chirality structure of the interaction and the sign of its interference with the standard model Drell--Yan background. The observed limits are in the range from 6.9 to 9.0\TeV for the \bbll signal based on different chirality and interference assumptions. These results represent the first LHC constraints on the \bbll model, significantly extending the sensitivity beyond the limits previously set at LEP. In the \bsll model, lower limits on the ratio of the energy scale of new physics to the coupling (\lamdagstar) range from 1.9 to 2.2\TeV depending on the channel and 2.4\TeV for the combination of all the channels. Additionally, the $\Pe\Pe$ and $\PGm\PGm$ invariant mass spectra are compared at the \TeVns scale. No significant deviation in the lepton flavor ratio from the standard model expectation is observed.

\begin{acknowledgments}
We congratulate our colleagues in the CERN accelerator departments for the excellent performance of the LHC and thank the technical and administrative staffs at CERN and at other CMS institutes for their contributions to the success of the CMS effort. In addition, we gratefully acknowledge the computing centers and personnel of the Worldwide LHC Computing Grid and other centers for delivering so effectively the computing infrastructure essential to our analyses. Finally, we acknowledge the enduring support for the construction and operation of the LHC, the CMS detector, and the supporting computing infrastructure provided by the following funding agencies: SC (Armenia), BMBWF and FWF (Austria); FNRS and FWO (Belgium); CNPq, CAPES, FAPERJ, FAPERGS, and FAPESP (Brazil); MES and BNSF (Bulgaria); CERN; CAS, MoST, and NSFC (China); MINCIENCIAS (Colombia); MSES and CSF (Croatia); RIF (Cyprus); SENESCYT (Ecuador); ERC PRG, RVTT3 and MoER TK202 (Estonia); Academy of Finland, MEC, and HIP (Finland); CEA and CNRS/IN2P3 (France); SRNSF (Georgia); BMBF, DFG, and HGF (Germany); GSRI (Greece); NKFIH (Hungary); DAE and DST (India); IPM (Iran); SFI (Ireland); INFN (Italy); MSIP and NRF (Republic of Korea); MES (Latvia); LMTLT (Lithuania); MOE and UM (Malaysia); BUAP, CINVESTAV, CONACYT, LNS, SEP, and UASLP-FAI (Mexico); MOS (Montenegro); MBIE (New Zealand); PAEC (Pakistan); MES and NSC (Poland); FCT (Portugal); MESTD (Serbia); MICIU/AEI and PCTI (Spain); MOSTR (Sri Lanka); Swiss Funding Agencies (Switzerland); MST (Taipei); MHESI and NSTDA (Thailand); TUBITAK and TENMAK (Turkey); NASU (Ukraine); STFC (United Kingdom); DOE and NSF (USA).

\hyphenation{Rachada-pisek} Individuals have received support from the Marie-Curie program and the European Research Council and Horizon 2020 Grant, contract Nos.\ 675440, 724704, 752730, 758316, 765710, 824093, 101115353, 101002207, 101001205, and COST Action CA16108 (European Union); the Leventis Foundation; the Alfred P.\ Sloan Foundation; the Alexander von Humboldt Foundation; the Science Committee, project no. 22rl-037 (Armenia); the Fonds pour la Formation \`a la Recherche dans l'Industrie et dans l'Agriculture (FRIA-Belgium); the Beijing Municipal Science \& Technology Commission, No. Z191100007219010, the Fundamental Research Funds for the Central Universities, the Ministry of Science and Technology of China under Grant No. 2023YFA1605804, and the Natural Science Foundation of China under Grant No. 12061141002 (China); the Ministry of Education, Youth and Sports (MEYS) of the Czech Republic; the Shota Rustaveli National Science Foundation, grant FR-22-985 (Georgia); the Deutsche Forschungsgemeinschaft (DFG), among others, under Germany's Excellence Strategy -- EXC 2121 ``Quantum Universe" -- 390833306, and under project number 400140256 - GRK2497; the Hellenic Foundation for Research and Innovation (HFRI), Project Number 2288 (Greece); the Hungarian Academy of Sciences, the New National Excellence Program - \'UNKP, the NKFIH research grants K 131991, K 133046, K 138136, K 143460, K 143477, K 146913, K 146914, K 147048, 2020-2.2.1-ED-2021-00181, TKP2021-NKTA-64, and 2021-4.1.2-NEMZ\_KI-2024-00036 (Hungary); the Council of Science and Industrial Research, India; ICSC -- National Research Center for High Performance Computing, Big Data and Quantum Computing and FAIR -- Future Artificial Intelligence Research, funded by the NextGenerationEU program (Italy); the Latvian Council of Science; the Ministry of Education and Science, project no. 2022/WK/14, and the National Science Center, contracts Opus 2021/41/B/ST2/01369 and 2021/43/B/ST2/01552 (Poland); the Funda\c{c}\~ao para a Ci\^encia e a Tecnologia, grant CEECIND/01334/2018 (Portugal); the National Priorities Research Program by Qatar National Research Fund; MICIU/AEI/10.13039/501100011033, ERDF/EU, "European Union NextGenerationEU/PRTR", and Programa Severo Ochoa del Principado de Asturias (Spain); the Chulalongkorn Academic into Its 2nd Century Project Advancement Project, and the National Science, Research and Innovation Fund via the Program Management Unit for Human Resources \& Institutional Development, Research and Innovation, grant B39G680009 (Thailand); the Kavli Foundation; the Nvidia Corporation; the SuperMicro Corporation; the Welch Foundation, contract C-1845; and the Weston Havens Foundation (USA).  
\end{acknowledgments}

\bibliography{auto_generated} 

\cleardoublepage \appendix\section{The CMS Collaboration \label{app:collab}}\begin{sloppypar}\hyphenpenalty=5000\widowpenalty=500\clubpenalty=5000\cmsinstitute{Yerevan Physics Institute, Yerevan, Armenia}
{\tolerance=6000
V.~Chekhovsky, A.~Hayrapetyan, V.~Makarenko\cmsorcid{0000-0002-8406-8605}, A.~Tumasyan\cmsAuthorMark{1}\cmsorcid{0009-0000-0684-6742}
\par}
\cmsinstitute{Institut f\"{u}r Hochenergiephysik, Vienna, Austria}
{\tolerance=6000
W.~Adam\cmsorcid{0000-0001-9099-4341}, J.W.~Andrejkovic, L.~Benato\cmsorcid{0000-0001-5135-7489}, T.~Bergauer\cmsorcid{0000-0002-5786-0293}, K.~Damanakis\cmsorcid{0000-0001-5389-2872}, M.~Dragicevic\cmsorcid{0000-0003-1967-6783}, C.~Giordano, P.S.~Hussain\cmsorcid{0000-0002-4825-5278}, M.~Jeitler\cmsAuthorMark{2}\cmsorcid{0000-0002-5141-9560}, N.~Krammer\cmsorcid{0000-0002-0548-0985}, A.~Li\cmsorcid{0000-0002-4547-116X}, D.~Liko\cmsorcid{0000-0002-3380-473X}, I.~Mikulec\cmsorcid{0000-0003-0385-2746}, J.~Schieck\cmsAuthorMark{2}\cmsorcid{0000-0002-1058-8093}, R.~Sch\"{o}fbeck\cmsAuthorMark{2}\cmsorcid{0000-0002-2332-8784}, D.~Schwarz\cmsorcid{0000-0002-3821-7331}, M.~Sonawane\cmsorcid{0000-0003-0510-7010}, W.~Waltenberger\cmsorcid{0000-0002-6215-7228}, C.-E.~Wulz\cmsAuthorMark{2}\cmsorcid{0000-0001-9226-5812}
% ...
\par}
\vskip\cmsinstskip
\dag:~Deceased\\
$^{1}$Also at Yerevan State University, Yerevan, Armenia\\
$^{2}$Also at TU Wien, Vienna, Austria\\
% ...
\end{sloppypar}
\end{document}